\documentclass[%
 reprint,
 amsmath,amssymb,
 aps,
]{revtex4-2}
\usepackage{physics}
\usepackage{graphicx}
\usepackage{dcolumn}

\usepackage{subfigure}
\usepackage{dsfont}

\usepackage[english]{babel}
\usepackage[utf8]{inputenc}
\usepackage[T1]{fontenc}
\usepackage{mathrsfs}
\usepackage{bbm}
\usepackage{bm}
\usepackage{enumerate}
\usepackage[dvipsnames]{xcolor}
\usepackage{float}
\usepackage{subdepth}
\usepackage{fnpct}
\usepackage{booktabs}
\usepackage{multirow}
\usepackage[math]{cellspace}
\usepackage{lipsum}
\usepackage{comment}
\usepackage{soul, ulem}

\usepackage[colorlinks,
linkcolor=BrickRed,
citecolor=MidnightBlue,
urlcolor=MidnightBlue,
bookmarks=true,        
bookmarksopen=true,    
bookmarksnumbered=true,
]{hyperref}

\newcounter{pr}

\begin{document}

\preprint{APS/123-QED}

\title{
Quantum Liouvillian Tomography}
 
\author{Diogo Aguiar}
\affiliation{CeFEMA-LaPMET, Physics Department, Instituto Superior Técnico, Universidade de Lisboa, 1049-001 Lisboa, Portugal}

\author{Kristian Wold}
\affiliation{Department of Computer Science, OsloMet – Oslo Metropolitan University, N-0130 Oslo, Norway}

\author{Sergey Denisov}
\affiliation{Department of Computer Science, OsloMet – Oslo Metropolitan University, N-0130 Oslo, Norway}

\author{Pedro Ribeiro}\email{ribeiro.pedro@tecnico.ulisboa.pt}
\affiliation{CeFEMA-LaPMET, Physics Department, Instituto Superior Técnico, Universidade de Lisboa, 1049-001 Lisboa, Portugal}
\affiliation{Beijing Computational Science Research Center, Beijing 100193, China}


\date{\today}

\begin{abstract}

Characterization of near-term quantum computing platforms requires the ability to capture and quantify dissipative effects. This is an inherently challenging task, as these effects are multifaceted, spanning a broad spectrum from Markovian to strongly non-Markovian dynamics. 
We introduce Quantum Liouvillian Tomography (QLT), a protocol 
to capture and quantify non-Markovian 
effects in time-continuous quantum dynamics.
The protocol leverages gradient-based quantum process tomography to reconstruct dynamical maps and utilizes regression over the derivatives of Pauli string probability distributions to extract the Liouvillian governing the dynamics.
We benchmark the protocol using synthetic data and quantify its accuracy in recovering Hamiltonians, jump operators, and dissipation rates for two-qubit systems.

Finally, we apply QLT to analyze the evolution of an idling two-qubit system implemented on 
a superconducting quantum platform to extract characteristics of Hamiltonian and dissipative components and, as a result, detect inherently non-Markovian dynamics. 
Our work introduces the first protocol capable of 
retrieving generators of generic open quantum evolution from experimental data, thus enabling more precise characterization of many-body non-Markovian effects in near-term quantum computing platforms.
\end{abstract}

\maketitle


\section{\label{sec:level1}Introduction} 

The rapid advancement of quantum technologies -- including quantum computing, communication, and metrology -- relies  on the precise control of quantum systems. As a result, the accurate characterization and validation of quantum processes has become a central challenge across these fields~\cite{eisert_quantum_2020,blume2025}.

In particular, methods for identifying and characterizing errors in quantum  platforms~ \cite{eisert_quantum_2020, gebhart_learning_2023}
vary significantly in terms of their capabilities, computational costs, and the assumptions made about the evolution of the systems and factors affecting it. Optimizing all three aspects simultaneously is challenging, and most protocols usually prioritize two aspects while compromising on the third. Randomized benchmarking~\cite{magesan_characterizing_2012}, for example, estimates the average error rates of quantum gates under minimal assumptions about the system dynamics, while combining low experimental overhead with the ability to assign technology-independent error metrics to multi-qubit gate-based platforms.

Quantum Process Tomography (QPT) ~\cite{mohseni_quantum_2008,kiktenko_estimating_2020} provides a framework for reconstructing an unknown quantum operation (also addressed as "map"~\cite{watrous2018} or "channel"~\cite{holevo2019}), thus offering insight into how current Noisy-Intermediate Scale quantum (NISQ) devices~\cite{preskill_quantum_2018} process information. 
QPT is expected to play an important role in hardware validation, quantum error correction, and the optimization of quantum algorithms, by enabling comprehensive characterization of noise and gate errors. However, while providing insights into quantum processes, QPT techniques usually incur substantial computational cost on both the classical and quantum levels. Potentially, this cost can be mitigated by making simplifying assumptions, such as, e.g., the assumption of rank-deficient states~ \cite{eisert_quantum_2020, gebhart_learning_2023}.

Given the challenges of QPT, several alternative approaches have been developed to improve efficiency and scalability. Methods such as compressive sensing~\cite{Foucart2013}, when combined with machine learning~\cite{gebhart2023}, as well as tensor network techniques~\cite{Torlai2020}, help to exploit the sparsity of quantum processes and reduce the number of required measurements while maintaining needed retrieval accuracy~\cite{Shabani2011, Torlai2023}. In addition, Fourier Quantum Process Tomography (FQPT)~\cite{DiColandrea2024} and  classical shadow techniques~\cite{Levy2024} help to reduce  measurement overhead. Although these techniques improve scalability, their performance nevertheless  remains limited when they applied to many-body quantum systems and non-unitary processes. 

Most importantly, QPT is fundamentally limited in its ability to independently characterize unitary and non-unitary errors, as a clear conceptual separation between them is not possible within the framework of quantum processes (or maps, or channels). Recently, significant progress has been made in reconstructing generators of open quantum dynamics, aiming to recover the underlying dynamical equations that govern time-continuous evolution, rather than descriptions in the form of operations (maps). The proposed \textit{Lindblad} tomography methods~\cite{onorati_fitting_2023,samach_lindblad_2022,olsacher_hamiltonian_2024,howard_quantum_2006,pastori_characterization_2022} assume that the generators have the celebrated Gorini–Kossakowski–Sudarshan–Lindblad (GKSL) form~\cite{GKS,L}, and enable  direct extraction of these generators from experimental data, thereby allowing for  independent quantitative analysis of unitary and non-unitary mechanisms.
It has also been demonstrated that many-body Lindbladians can be efficiently reconstructed from local measurements, provided that physically motivated constraints, such as locality and finite-range interactions, hold~\cite{StilckFranca2024}. This potentially opens a pathway toward scalable, generator-based tomography for multi-qubit quantum processors.

However, the accurate retrieval of quantum generators necessitates a proper treatment of non-Markovian effects~\cite{rivas2014,Breuer2016,Chruscinski2022}, which include, e.g., memory effects and information backflows from the environment to the system. There are recent theoretical and experimental developments demonstrating that multi-time process methods reveal substantial non-Markovian effects in superconducting quantum platforms~\cite{white2020,white_non-markovian_2022}. This  "process tensor
framework"~\cite{pollock2018nonmarkovian} enables capturing long-time correlations through a chain of chronologically ordered maps, retrieved from the system dynamics, yet it inherently precludes  decomposition of the system dynamics into unitary and non-unitary components.  On the other hand, existing Lindblad tomography schemes~\cite{onorati_fitting_2023,samach_lindblad_2022,olsacher_hamiltonian_2024,howard_quantum_2006,pastori_characterization_2022} naturally assume Markovian quantum evolution and are therefore incapable of capturing temporal correlations and structured noise.

In this work, we propose Quantum Liouvillian Tomography (QLT), a protocol that generalizes existing Lindblad tomography methods~\cite{onorati_fitting_2023,samach_lindblad_2022,olsacher_hamiltonian_2024,howard_quantum_2006,pastori_characterization_2022,StilckFranca2024} to the case of non-Markovian evolution. By using synthetic and experimental data, collected on the superconducting quantum platform~\cite{helmiQC}, we not only demonstrate the ability of the protocol to capture non-Markovian effects, but also show that it can extract key characteristics of both the Hamiltonian and dissipative components of the underlying dynamics.

The paper is organized as follows. In Section~\ref{sec:First}, we introduce the QLT protocol for a time-resolved 
Liouvillian retrieval, based on a regression procedure. The application of QLT is presented in Section~ \ref{sec:Third}, where we first benchmark the protocol with synthetic data  and then use it to analyze  experimental data. The final conclusions and outlook are presented in Section~\ref{sec:Outlook}.

\section{Quantum Liouvillian Tomography: A protocol \label{sec:First}}

\subsection{General considerations on Quantum Liouvillians \label{sec:QL}}

We assume that the time-continuous evolution of a quantum system, initially uncorrelated with its environment,  can be described by a family of completely positive trace reserving (CPTP) maps~\cite{watrous2018} \(\Lambda_t\), acting on an initial state \(\rho_0\), $\rho(t) = \Lambda_t(\rho_0)$.    
This dynamics can be seen as generated by a time-local operator \(\mathcal{L}_t\), satisfying  
\begin{equation}
    \frac{d}{dt}\Lambda_t = \mathcal{L}_t \Lambda_t. \label{eq:LindGen}
\end{equation}  
The generator \(\mathcal{L}_t\) admits a generalized Lindblad-like form~\cite{hall_canonical_2014},  
\begin{align}
    \mathcal{L}_t(\cdot) = & -i\comm{H(t)}{\cdot}  \label{eq:LindbladForm} \\
    + & \sum_{\mu} \gamma_\mu (t) \left( J_\mu(t) \cdot J_\mu^\dagger(t) - \frac{1}{2}\acomm{J_\mu (t)J_\mu^\dagger(t)}{\cdot} \right), \nonumber
\end{align}  
where \( H(t) \) governs unitary evolution, while the (non-necessarily non-negative) dissipation rates, \(\gamma_\mu (t)\), and jump operators \(J_\mu (t)\) describe dissipation and decoherence. This decomposition is unique if: \((i)\) \(H\) is traceless, and \((ii)\) \(\{J_\mu\}\) are traceless and orthonormal under the Hilbert-Schmidt inner product, $\langle A,B \rangle = \Tr{A B^\dagger}$~\cite{manzano_short_2020}.

For time-independent \(\mathcal{L}_t\) with \(\gamma_\mu \geq 0\), the standard Lindblad form is recovered. However, in general, time-dependent \(\gamma_\mu\) may take negative values for \(t > 0\), indicating non-Markovian dynamics~\cite{hall_canonical_2014}. This implies that the evolution from \(t_i\) to \(t_f > t_i\), given by  
\begin{equation}
    \Lambda_{t_f,t_i} = \Lambda_{t_f} \Lambda_{t_i}^{-1},  
\end{equation}  
is not necessarily CPTP, characterizing CP-indivisible dynamics. Such behavior can only arise in the presence of non-trivial system-environment correlations that require information backflow between the system and the environment. Here, we adopt CP-indivisibility as the definition of quantum non-Markovianity~\cite{breuer_non-markovian_2016}.  

The following sections outline a method to reconstruct the time-dependent Liouvillian \(\mathcal{L}_t\) from tomographic data.

\subsection{Reconstructing Maps and SPAM parameters \label{sec:PS}}
\begin{figure}
    \subfigure{\includegraphics[width=0.55\textwidth]{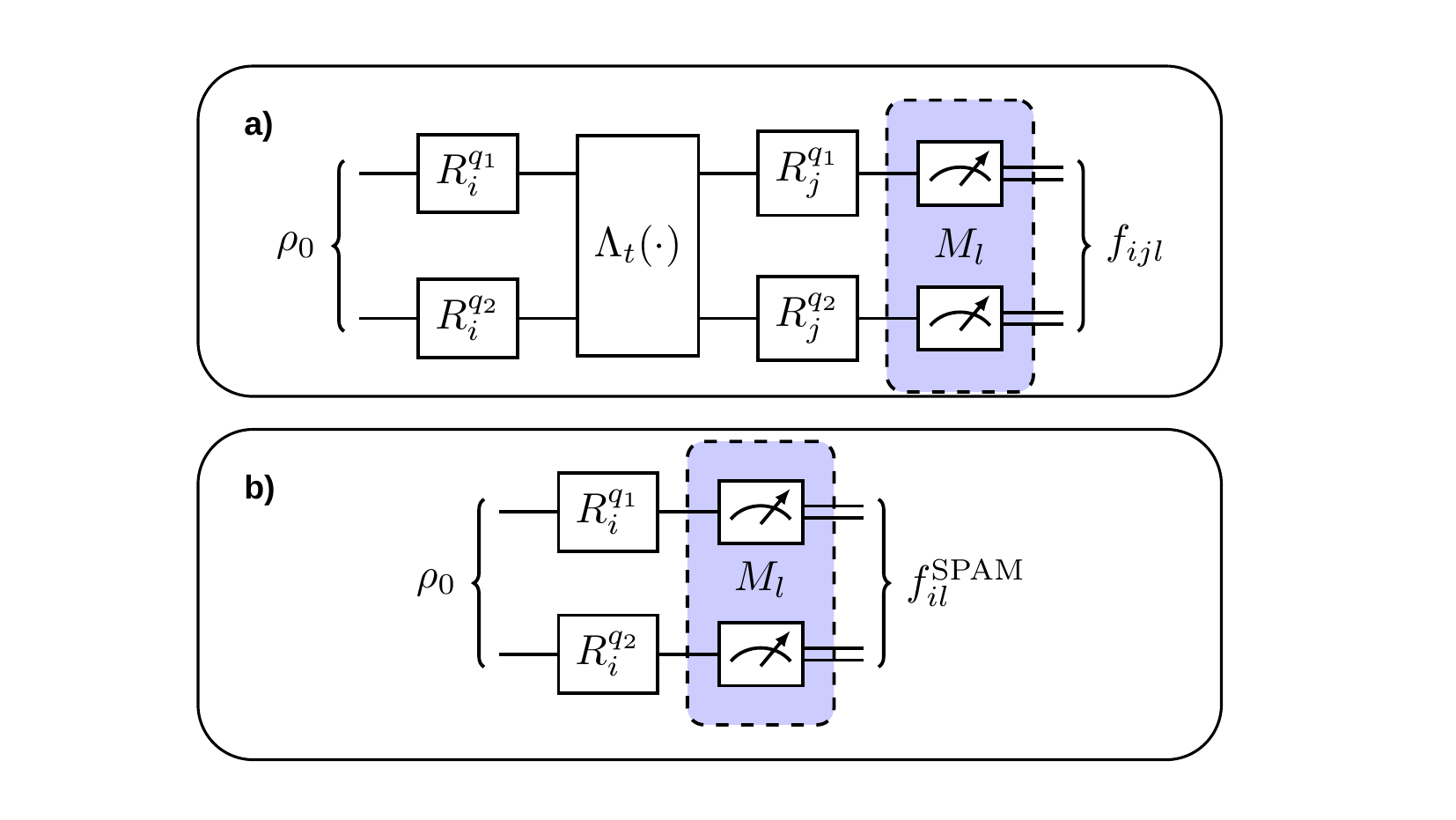}}  
    \caption{Diagram of Pauli string modes. a) circuits employed for process and Liouvillian tomography. b) circuits employed for SPAM tomography. (Diagram drawn with Quantizk \cite{kay2023tutorialquantikzpackage})}
    \label{fig:PauliCircuits}
\end{figure}
We consider a black-box circuit of $N_q$ qubits subject to an unknown time-dependent operation, $\Lambda_t$, which can be consistently replicated. The circuit is initialized in a fixed state $\rho_0$ and measured via a Positive Operator-Valued Measurement (POVM) $\{M_l > 0\}$ satisfying  
\begin{equation}
    \sum_l M_l = 1.
\end{equation}
The combination $(\rho_0, M_l)$ defines the state preparation and measurement (SPAM) parameters, which are assumed constant throughout the experiment.

To characterize $\Lambda_t$, we employ Pauli string tomography~\cite{samach_lindblad_2022, wold_universal_2024, roncallo2024pauli}, an efficient method for reconstructing quantum processes. This involves preparing a set of input states, $\rho_i = R_i \rho_0 R_{i}^\dagger$, where $R_i = \bigotimes_{q_i=1}^{N_q} R_i^{q_i}, $ consists of single-qubit rotations chosen from $ R_i^{q_i}\in \{R_Y(\pm\pi/2),R_X(\pm \pi/2),X,\mathds{1}\}$, and applying the unknown operation $\Lambda_t$. Before measurement, an additional set of rotations $R_j$ is applied to transform $\Lambda_t(\rho_i)$ into a suitable basis, leading to the final state:
\begin{equation}
    \rho_f = R_j \Lambda_t (R_i \rho_0 R_i^\dagger) R_j^\dagger.
\end{equation}
Each measurement produces an $N_q$-bit classical outcome, and the probability of obtaining a given result $l$ follows Born’s rule:
\begin{equation}
    p_{l|ij}(t) = \Tr{\Lambda_t(\rho_i) M_{lj}}, \quad M_{lj} = R_j^\dagger M_l R_j.
    \label{eq:PauliStringProb}
\end{equation}
The circuits implementing these operations are illustrated in Fig.~\ref{fig:PauliCircuits}.

In the ideal case where $\rho_0 = \ket{\boldsymbol{0}}\bra{\boldsymbol{0}}$ and the POVM corresponds to computational basis measurements, Pauli string tomography simplifies to preparing and measuring qubits in eigenstates of the Pauli operators $\sigma_{\alpha}$, with $\alpha = x,y,z$. However, real quantum hardware provides only classical measurement outcomes, and probabilities in Eq.~\eqref{eq:PauliStringProb} must be estimated from empirical frequencies. Given $N_s$ circuit executions (shots), the probability is approximated as $\tilde{p}_{l|ij} = f_{ijl} / N_s$, where $f_{ijl}$ is the observed frequency of outcome $l$. Statistical fluctuations introduce an uncertainty of order $\mathcal{O}(1/\sqrt{N_s})$, necessitating a trade-off between precision and execution time.

While there are $18^{N_q}$ possible Pauli strings, practical implementations use a reduced set $N_p \ll 18^{N_q}$ with minimal performance loss, mitigating the exponential scaling in system size~\cite{Shabani2011, vanDenBerg2023probabilistic}. For larger systems, sparsity assumptions on the Liouvillian operator can further enhance scalability, as demonstrated in recent probabilistic recovery methods~\cite{vanDenBerg2023probabilistic}. The following sections outline a protocol for reconstructing $\mathcal{L}_t$ from tomographic data.

The reconstruction of SPAM parameters $(\rho_0, M_l)$ and of the quantum channel $\Lambda_t$ can be achieved using established tomographic techniques~\cite{samach_lindblad_2022, wold_universal_2024}. In brief, both estimations require additional Pauli string measurements: SPAM characterization employs circuits depicted in Fig.~\ref{fig:PauliCircuits}-(b) to obtain probabilities $p_{l|i}^{\text{SPAM}}$, while quantum channel reconstruction utilizes circuits  Fig.~\ref{fig:PauliCircuits}-(a)  to estimate probabilities $p_{l|ij}(t)$. Detailed protocols used in the following and their performance benchmarks are provided in Appendices~\ref{sec:SPAMTom} and \ref{sec:MapTom}.

\subsection{Liouvillian estimation \label{sec:LTProtocol}}

To estimate the Liouvillian $\mathcal{L}_t$ governing a quantum system’s dynamics, we consider the time derivative of the Pauli string probabilities:
\begin{equation}
\frac{d}{dt} p_{l|ij}(t) = \mathrm{Tr}\left\{\mathcal{L}_t \Lambda_t(\rho_i) M_{lj}\right\},
\label{eq:DiffProb}
\end{equation}
where $p_{l|ij}(t)$ denotes the probability of obtaining outcome $l$ given input state $\rho_i$ and measurement $M_{lj}$ after process $\Lambda_t$. Retrieving $\mathcal{L}_t$ requires first:
(i) determining SPAM parameters $(\rho_0, M_l)$;
(ii) characterizing the quantum channel $\Lambda_t$;
(iii) estimating the time derivatives $\frac{d}{dt} \tilde{p}_{l|ij}(t)$. 
Steps (i) and (ii) are given in the previous section. For step (iii), in practice, time derivatives are approximated using finite difference methods $\frac{d}{dt} \tilde{p}_{l|ij}(t) \simeq  [\tilde{p}_{l|ij}(t+dt) -  \tilde{p}_{l|ij}(t)]/dt$, which additionally requires acquiring tomographic data for time $t+dt$. Here, selecting an appropriate time step $dt$ is crucial: a small $dt$ may lead to significant statistical noise, while a large $dt$ can introduce discretization errors in time-dependent Liouvillians. Techniques such as polynomial interpolation may be also used to enhance the robustness of these estimations \cite{StilckFranca2024}.

To retrieve the Liouvillian from Eq.~(\ref{eq:DiffProb}), having completed the steps (i-iii), we formulate an optimization problem:
\begin{equation}
\tilde{\theta} = \arg \min_{\theta} \frac{1}{2^{N_q} N_p} \sum_{i,j,l} \left[ \frac{d}{dt} \tilde{p}_{l|ij}(t) - \mathrm{Tr}\left\{\mathcal{L}(\theta) \Lambda_t(\rho_i) M_{lj}\right\} \right]^2,
\label{eq:LiouvillianMinimization}
\end{equation}
where $\theta = (h,K)$ parameterize the Liouvillian operator. 

This parametrization is more easily described in vectorized notation, with density matrices mapped into $d^2 = 4^{N_q}$ vectors and superoperators mapped into $d^2\times d^2$ matrices. 
\begin{align}
    &\mathcal{L}(\theta) = -i(H\otimes \mathds{1} - \mathds{1} \otimes H^T) +\nonumber\\
    &\sum_{\mu,\nu = 1}^{d^2-1}\gamma_{\mu\nu} \left[ G_\mu\otimes G_\nu^* -\frac{1}{2}\left(G_\nu^\dagger G_\mu\otimes \mathds{1} + \mathds{1}\otimes \left(G_\nu^\dagger G_\mu\right)^T \right) \right]\label{eq:LindParam},
\end{align}
where $\{G_\mu\}$ is an orthonormal basis of operators with respect to the Hilbert-Schmidt inner-product, and 
\begin{align}
    H = \sum_{\mu=1}^{d^2} h_\mu G_\mu, \,\,\, 
    \gamma = K + K^\dagger. \label{eq:GammaParam}
\end{align} 
Taking the operator basis to be hermitian, $G_i^\dagger = G_i$, 
one gets $h\in\mathbb{R}^{d^2-1}$ and $K\in\mathbb{C}^{(d^2-1)\times (d^2-1)}$. Specifically, in the following we take $G_\mu$ to be normalized products of Pauli matrices:
$G_\mu = \frac{1}{d}\otimes_{\mu_i}^{N_q}\sigma_{\mu_i},$ with $\mu_i = 0,x,y,z$ and $\sigma_0 = \mathds{1}$.

The optimization is done over a space of unconstrained parameters $\theta = (h,K)$. Eq.~(\ref{eq:LiouvillianMinimization}) is solved with recourse to the Adam optimizer \cite{kingma2017adammethodstochasticoptimization},
a gradient descent-like method of iteratively computing the MSE at points in $\theta$-space that align with the direction of decreasing gradient.

Once the optimal parameters are determined, the Liouvillian can be put in the canonical form of Eq.~(\ref{eq:LindGen}) by diagonalizing the matrix $\gamma$,
\begin{align}
    \gamma_{\delta\delta'} &= \sum_\mu \lambda_\mu j_{\mu\delta},  j_{\mu\delta'}^*, \label{eq:CanonicalJOpRates} 
\end{align}
and taking the eigenvalues as the dissipation rates $\gamma_\mu (t) = \lambda_\mu,$  and the normalized eigenvectors to encode the respective jump operators $ J_\mu (t)= \sum_\delta j_{\mu\delta} G_\delta$.

\subsection{Error estimation\label{sec:ErrorSources}}

To quantify uncertainties in Liouvillian tomography, we employ a bootstrap approach. Specifically, we repeatedly resample measured probabilities using the original shot counts, generating $M$ synthetic datasets. For each dataset, we reconstruct the SPAM parameters, quantum channel, and Liouvillian, following the procedure described above.

The statistical error in an observable $O$ derived from the Liouvillian (e.g., dissipative rates) is estimated as the standard deviation across bootstrap samples:
\begin{equation}
    \delta O = \sqrt{\frac{1}{M}\sum_{m=1}^{M}\left(O^{(m)} - \bar{O}\right)^2},
\end{equation}
where $\bar{O}$ denotes the estimate obtained with the original dataset. This procedure approximates the statistical uncertainty introduced by finite-shot noise propagation through all reconstruction steps. We emphasize that it does not account for systematic errors arising from experimental imperfections or model assumptions. 
To assess those, the only possibility is retrieving the Liouvillian with different values of $dt$, and verifying compatibility among the estimates.

\section{Results \label{sec:Third}}

\subsection{Benchmarking with simulated data}

To benchmark the QLT protocol, we simulate a two-qubit system ($N_q=2$) coupled to an environment of two qubits ($N_E=2$), itself interacting with a Markovian reservoir. The joint system-environment dynamics is governed by a static Liouvillian of the Lindblad from, composed of a Gaussian random Hamiltonian and randomly sampled dissipative jump operators. 
After tracing out the environment the evolution of the two-qubit system, $ \Lambda_t(\rho_0) = \text{Tr}_\text{E}[ e^{\mathcal{L}_\text{S+E} \,t } \rho_0 \otimes \rho_\text{E}] $, is generically non-Markovian~\cite{chruscinski_non-markovian_2010}. 
To test the performance of the procedure in a realistic setting, the SPAM parameters were sampled by randomly perturbing the ideal quantities, i.e., 
$\rho_0 = 0.9\ket{\boldsymbol{0}}\bra{\boldsymbol{0}} + 0.1\delta \rho,\, M_l = 0.8\ket{l}\bra{l} + 0.2\delta M_l$ for $1 \leq l \leq 15$, as detailed in Appendix \ref{sec:SPAMTom}. 
The estimated probabilities, $\tilde p_{l|ij}(t )$,  are generated for the full set of two-qubit Pauli Strings by sampling the exact probabilities with $N_s = 10^4$ shots.
The error estimation procedure of section \ref{sec:ErrorSources} was applied with $M = 6$.

To assess the accuracy of the retrieval procedure we compute the fidelity between the original ($\Lambda, \rho_0, \{M_l\}$) and retrieved quantities ($\tilde\Lambda, \tilde\rho_0, \{\tilde{M}_l\}$), defined as \cite{watrous_theory_2018,wold_universal_2024}:
\begin{subequations}
\begin{align}
    \mathcal{F}(\Lambda,\tilde{\Lambda}) & = \Tr{{\sqrt{\sqrt{\Phi_\Lambda}\Phi_{\Tilde{\Lambda}}\sqrt{\Phi_\Lambda}}}}^2,\\
    \mathcal{F}(\rho_0,\tilde{\rho}_0) &= \Tr{\sqrt{\sqrt{\rho_0} \tilde{\rho}_0\sqrt{\rho_0}}}^2,\\
    \mathcal{F}(\{M_l\},\{\tilde{M}_l\}) &= \frac{1}{d^2}\left(\sum_l \Tr{\sqrt{M_l \tilde{M}_l}}\right)^2,
\end{align}
\label{eq:Fidelity}
\end{subequations}
where $\Phi_\Lambda$ is the Choi matrix representation of $\Lambda$ \cite{watrous_theory_2018}.

The Liouvillian tomography protocol was applied for $t  \in \{0.1, 0.2, 0.3, 0.4\}$.
The results are summarized in Tab. \ref{table:ResultsLT}.

\begin{table}
  \centering
    \begin{tabular}{c|c|c|c|c}
    $t$  & $\mathcal{F}^{\rho_0}$ & $\mathcal{F}^M$ & $\mathcal{F}^\Lambda$ \\
    \hline
    0.1 & 0.980 & 0.989 & 0.967\\
    0.2 & 0.984 & 0.991 & 0.983\\ 
    0.3 & 0.967 & 0.968 & 0.972\\ 
    0.4 & 0.973 & 0.970 & 0.992\\ 
    \end{tabular}
  \caption{\label{table:ResultsLT}
  Benchmark results of SPAM and CPTP map reconstruction: the target quantities are compared to those retrieved with the methods of Appendices \ref{sec:SPAMTom} and \ref{sec:MapTom} for each evolution time by computing the fidelity measures of Eq. \ref{eq:Fidelity}.   
  }
\end{table}

The derivative of Pauli string distributions were computed at order $dt^2$ using:
$\frac{d}{dt}\tilde{p}_{l|ij}(t) = \frac{1}{2dt}(\tilde{p}_{l|ij}(t+dt) - \tilde{p}_{l|ij}(t -dt))$, with the values of $dt$  chosen such that the average of the estimated finite difference is larger than the error introduced by the finite number of measurement shots, i.e. $\text{avg}_{ijl}|\tilde{p}_{l|ij}(t+dt) - \tilde{p}_{l|ij}(t-dt)| \approx 0.02 > \frac{1}{\sqrt{N_s}}$.

\begin{figure*}
    \centering
    \includegraphics[width=\textwidth]{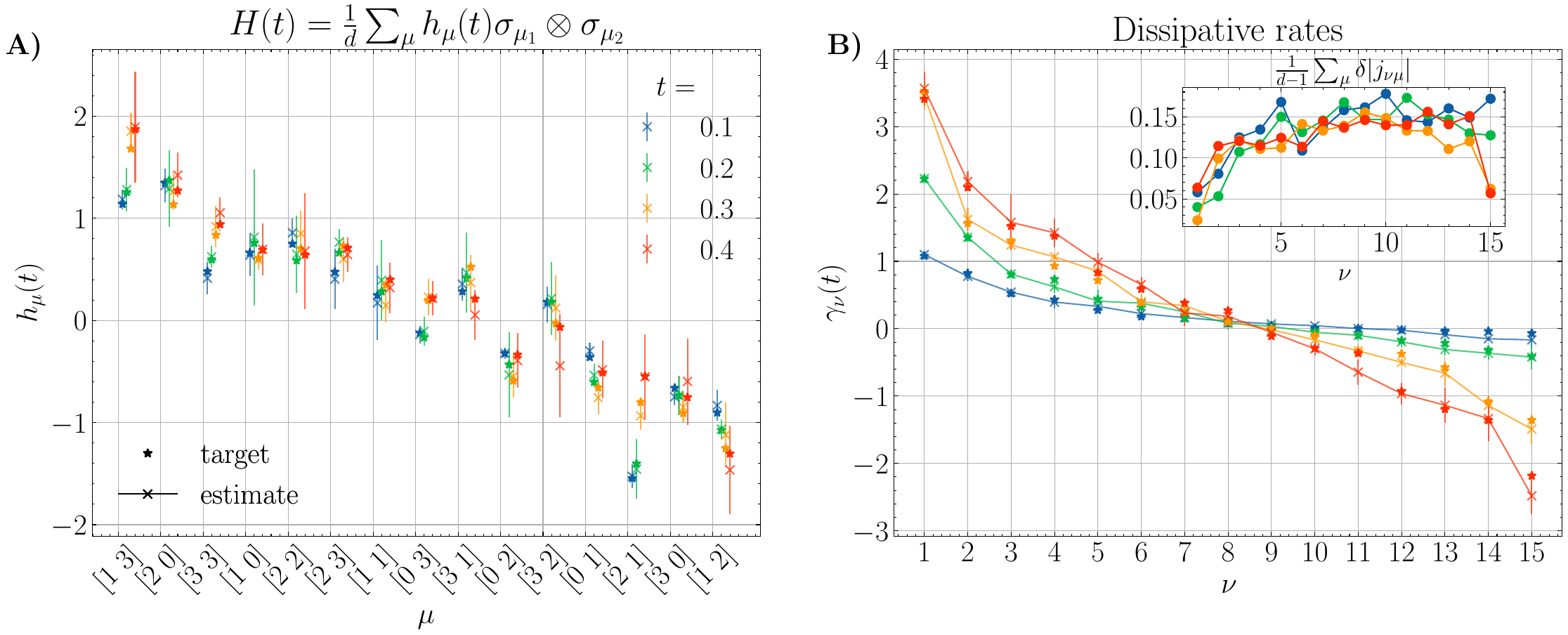}\\
    \includegraphics[width=\textwidth]{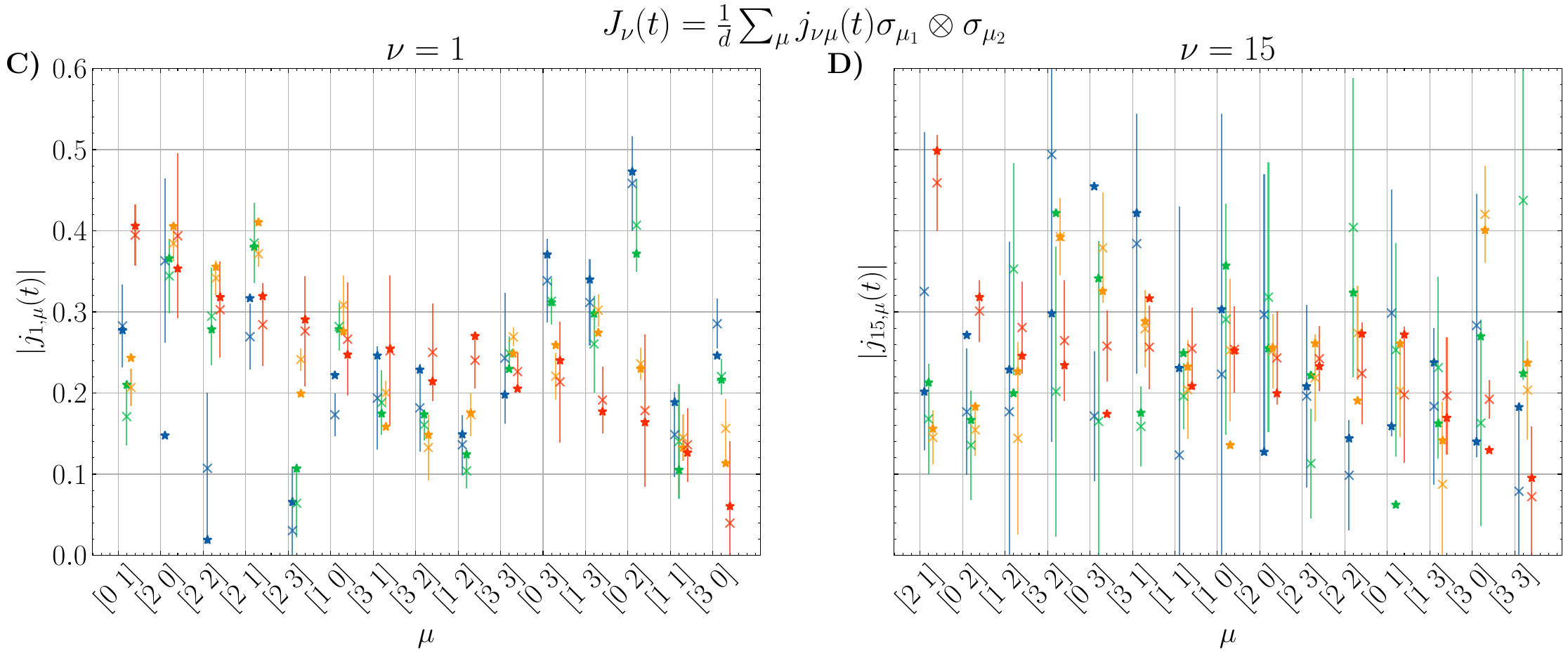}
    \caption{\label{fig:LT-Hamil} 
    QLT benchmark with simulated data for a generic non-Markovian evolution of a two qubits. Liouvillians are retrieved for $t \in \{0.1, 0.2, 0.3, 0.4\}$.
    QLT retrieved results are denoted by a cross, whereas the target values of the ground-truth Liouvillian are represented by a star. 
    (a) Components of the Hamiltonian, $h_\mu(t)$, decomposed over a Pauli base and ordered by decreasing magnitude. 
   (b) Dissipative rates $\gamma_\nu(t)$ in decreasing order. Inset: average estimated statistical error of each jump operator obtained according to Section~\ref{sec:ErrorSources} with $M = 6$.
   (c), (d) Components of the jump operators, $j_{\nu,\mu}(t)$, in the Pauli base, presented in decreasing order of $j_{\nu\mu}(t=0.4)$ for $\nu = 1,15$. 
    }
\end{figure*}

\begin{figure*}[tp]
    \includegraphics[width=\textwidth]{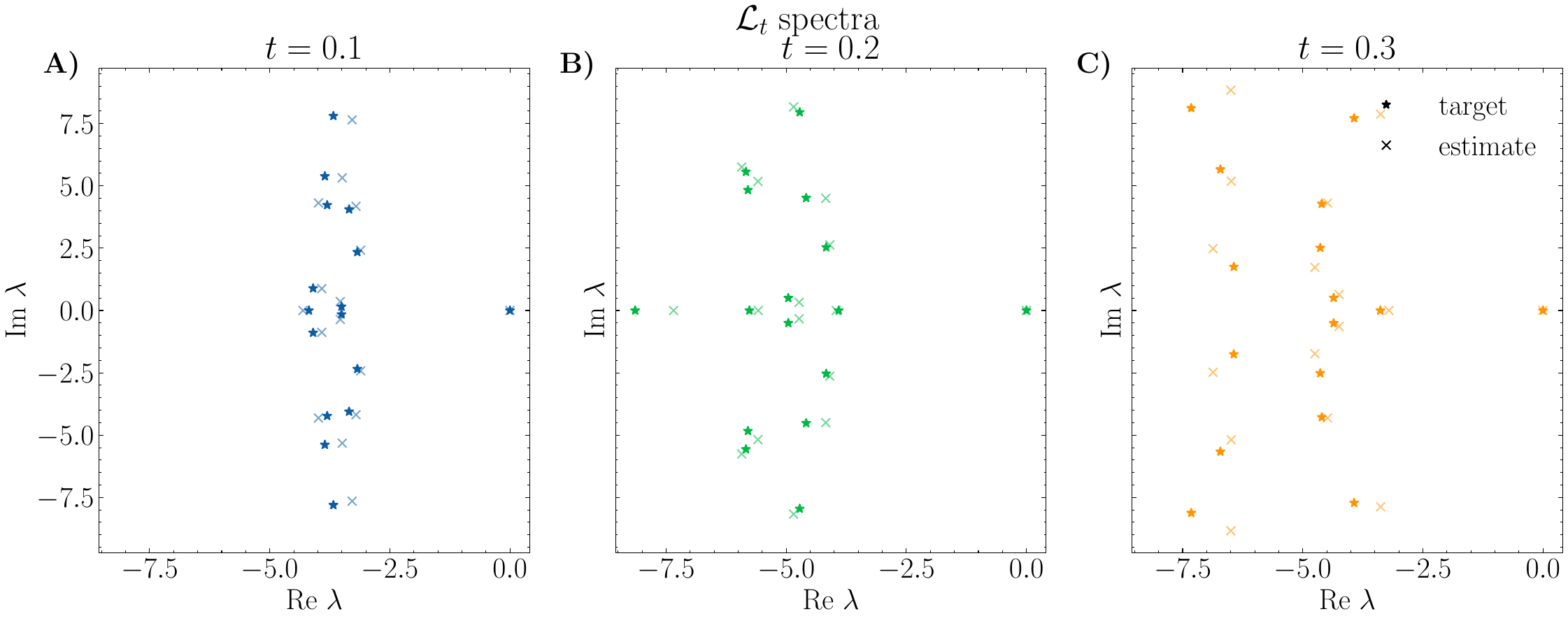}
    \caption{
    Spectral benchmarks. The spectrum of the target Liouvillian (stars) is compared with the spectrum of the estimated one (crosses) for different retrieval times.   
    }
    \label{fig:LT-Spectra}
\end{figure*}

The results are presented in Fig.~\ref{fig:LT-Hamil}.
Fig.~\ref{fig:LT-Hamil}-(a) depicts the components $h_\mu(t)$ of the Hamilonian decomposed in the Pauli basis, $\sigma_{\mu_1}\otimes\sigma_{\mu_2}$ for different times, presented in decreasing order of $h_\mu(t=0.4)$. The target values are compared with the estimations within the statistical error.

Fig.~\ref{fig:LT-Hamil}-(b) shows the dissipative rates $\gamma_\nu(t)$ in decreasing order. The inset displays the statistical error of the jump operators averaged over the components of each operator. 
Figs.~\ref{fig:LT-Hamil}-(c) and (d) show the components of the jump operators, $j_{\nu,\mu}(t)$ decomposed in the Pauli basis, presented in decreasing order of $j_{\nu\mu}(t=0.4)$ for $\nu = 1,15$. 

The retrieved Hamiltonians and dissipative rates reproduce accurately the target values, within the estimated errors. Moreover, non-Markovian behavior is correctly identified by the negative dissipative rates, even when these are of small magnitude.

On the other hand, the jump operators show stronger sensitivity to errors. Here, deviations from the target values exceeding the estimated error are more prevalent. Moreover, as the magnitude of the corresponding dissipative rate decreases, the estimates become increasingly noisy and the finite discretization introduces substantial systematic errors, that the already sizable error bars may still not account for. 
For instance, at $t = 0.1$, $\gamma_{15} \approx 0$ and the corresponding $j_{15\mu}$ is extremely noisy. However, as time increases $j_{15\mu}$ is estimated with increased precision. 
Such limitation are expected, as the jump operators associated to lower dissipative rates will have a lesser impact in the overall Liouvillian structure. 

Despite these discrepancies in the jump operators,  the comparison between the spectral of the original and retrieved operators, presented in Fig. \ref{fig:LT-Spectra}, is remarkable. Overall the spectral features of the target Liouvillian are reproduced and the eigenvalues of the retrieved and original operators can always be put in one-to-one correspondence. However, the quantitative agreement worsens for decreasing values of the $\text{Re}\lambda$, likely due to the fact that the respective eigenmodes decay faster and are harder to fit accurately specially for large times.

\subsection{Experimental results \label{sec:Fourth}}

In this section, we present results obtained using Helmi, a five-qubit quantum processor operated by the VTT Technical Research Center of Finland. By employing Liouvillian tomography, we characterized the dynamics of two qubits under idle conditions, i.e., when no explicit quantum operations are applied. Given that Helmi lacks a native idling gate, we devised an effective idle period by sequentially applying virtual $Z$ gates \cite{mckay_efficient_2017} on an auxiliary qubit. This approach allowed us to simulate an idle evolution interval corresponding approximately to integer multiples of the single-qubit gate duration, $T \approx 120,\text{ns}$.

Liouvillian generators were reconstructed for three idle durations, with normalized times $\tau \in \{3, 10, 20\} \times T $ . For each duration, the SPAM parameters and the corresponding CPTP map were estimated. Derivatives of the Pauli string distributions were estimated using $N_s = 2^{10}$  measurement shots and two discretization steps $dt = T $ and $dt = 2 T$.

\begin{figure*}
    \begin{center}
        \includegraphics[width=\textwidth]{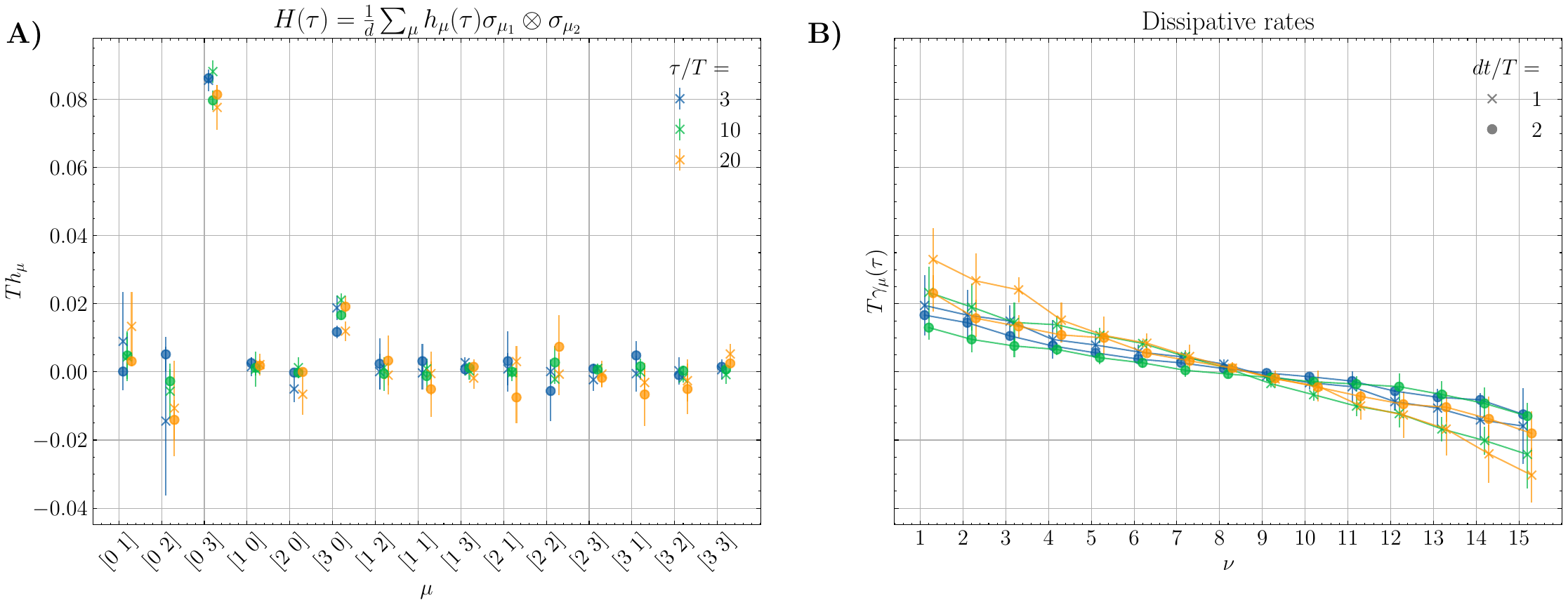}
    \end{center}
    \caption{\label{fig:HelmiHamilRates}
    Experimental results from Helmi during idle two-qubit dynamics. 
    The QLT procedure was applied for idling times $\tau \in \{3T, 10 T, 20T \}$ and two discretization steps $dt = T $ (crosses) and $dt = 2T$ (dots), and for $N_s = 2^{10}$ shots.
    (a) Components of the Hamiltonian, $h_\mu(\tau)$, decomposed over a Pauli basis. 
    (b) Dissipative rates $\gamma_\nu(\tau)$ in decreasing order. The error bars estimated by the procedure of Section \ref{sec:ErrorSources} with $M = 6$.
    }
\end{figure*}

The corresponding Hamiltonian and dissipative rates are depicted in Fig.~\ref{fig:HelmiHamilRates}.
For dissipative rates the error bars are substantially larger than those in the synthetic benchmarks. 

To access the performance of each stage in the Quantum Liouvillian tomography pipeline, we employ the R2 metric, defined as
\begin{equation}
    \text{R2} = 1 - \frac{\sum_{ijl}(y_{l|ij} - \tilde{y}_{l|ij})^2}{\sum_{ijl}(y_{l|ij} - \bar{y})^2},\quad \text{with } \bar{y} = \sum_{ijl}y_{l|ij},
\end{equation}
that quantifies the proportion of variance explained by the model, where $y_{l|ij}$ represents experimental data ($p_{l|ij}$ $d p_{l|ij}/dt$ ) and $\tilde{y}_{l|ij}$ the corresponding model predictions, i.e. either the probabilities or their derivatives computed with Born's rule or Eq.~(\ref{eq:DiffProb}), respectively. An R2 value approaching 1 indicates strong agreement.
Results are summarized in Table~\ref{table:ResultsHelmi}.
Note that R2 coefficients effectively measure how much better the estimated quantities explain the underlying data compared with the prediction of its mean value therefore providing a normalized metric to assess the quality of each minimization procedure.

\begin{table}
  \centering
    \begin{tabular}{c|c|c|c|c}
    $\tau$ & R2-SPAM  & R2-CPTP map & R2-QLT $dt = T$ & R2-QLT $dt = 2T$\\
    \hline
    3 & 0.996 & 0.979 & 0.855 & 0.945\\
    10 & 0.997 & 0.987 & 0.811 & 0.922 \\ 
    20 & 0.998 & 0.981 & 0.609 & 0.881 \\ 
    \end{tabular}
  \caption{\label{table:ResultsHelmi} Experimental results from Helmi: R2 values for minimization performed over Helmi data.}
\end{table}

High R2 values were obtained for both SPAM and CPTP estimation, while somewhat lower values were observed for Liouvillian tomography. This drop reflects the cumulative impact of statistical noise and imprecise prior estimates. Notably, the use of a coarser time grid ($dt = 2T$) improves the R2 coefficients, mitigating the effect of noise in the derivative estimates. Despite the degradation at $\tau = 20T$ for $dt = T$, all other configurations yield R2 values above $0.81$, indicating a reliable reconstruction of the Pauli derivative data.
Additionally, note that while inaccurate prior estimates, $(\tilde{\rho}_0, \{\tilde{M}_l\}, \tilde{\Lambda}_\tau)$, may decrease the R2 coefficient, the error estimation procedure already accounts for these factors.

The estimated Hamiltonian remains largely constant across the different idle durations and is dominated by single-qubit terms, particularly $\mathds{1}\otimes \sigma_z$ and  $\sigma_z \otimes \mathds{1}$,  consistent with relative phase accumulation due to energy splittings between $\ket{0}$ and  $\ket{1}$. In contrast, the dissipative rates exhibit greater sensitivity to the choice of discretization step, especially at longer idle times ($\tau = 10T$ and $20T$), suggesting non-negligible discretization errors. Nevertheless, negative dissipative rates persist across both grid sizes, signaling the presence of non-Markovian effects. The associated jump operators, however, show large variances and preclude definite physical interpretation of specific decoherence channels.

\begin{figure*}[tp]
    \includegraphics[width=\textwidth]{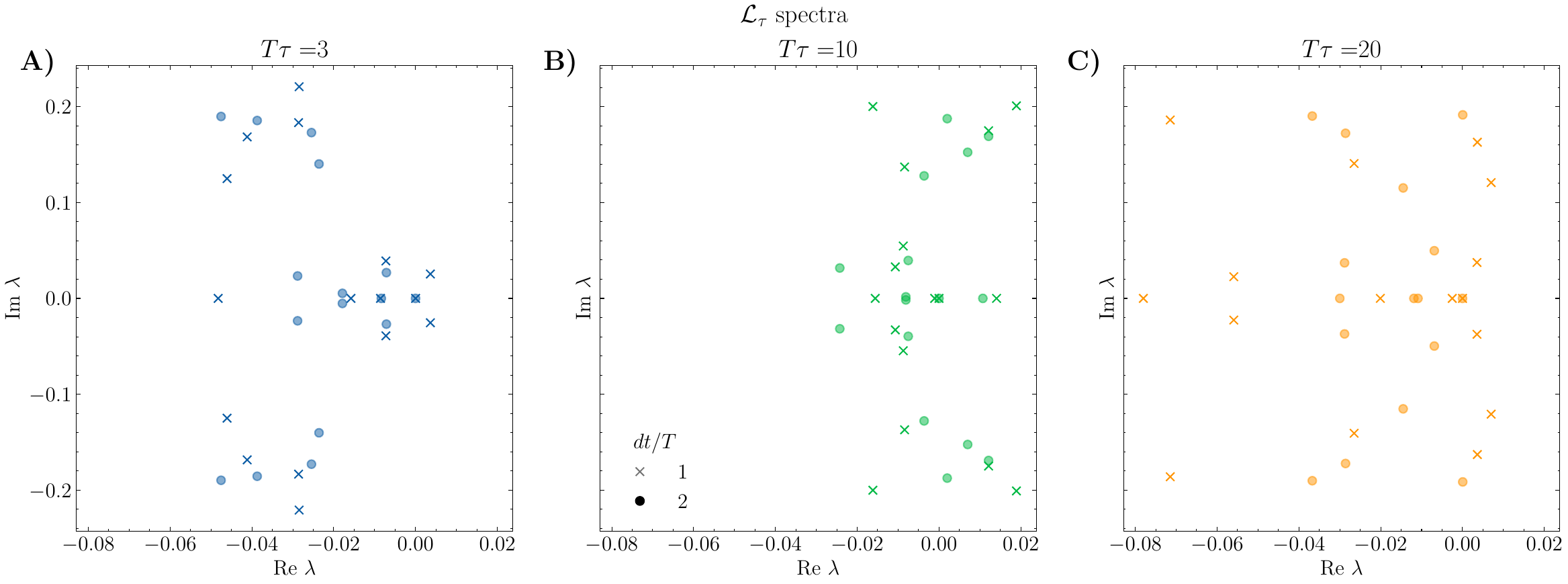}
    \caption{Experimental results for the spectrum from Helmi. 
    For each retrieval time, the spectra of the retrieved Liouvillians is shown for idling step $dt = T$ (crosses) and $dt = 2T$ (dots).
    }
    \label{fig:Helmi-Spectra}
\end{figure*}

The corresponding Liouvillian spectra are shown in Fig.~\ref{fig:Helmi-Spectra}, highlighting the differences induced by varying the discretization step. As expected, there are discrepancies at the level of dissipative rates and jump operators are reflected on the spectra with the eigenvalues estimated with each grid size differing substantially. 

These experimental results deviate substantially from simulations with synthetic data, reflecting the challenging experimental conditions intrinsic to the Helmi platform. We anticipate that access to pulse-level control, available on many current quantum processors \cite{alexander_qiskit_2020}, would enhance the precision and reliability of the method. While the uncertainty in the estimation of jump operators—and, to a lesser extent, dissipative rates—may suggest that full Liouvillian tomography was not achieved in this instance, our results demonstrate the method’s capability to extract effective Hamiltonians and identify non-Markovian behavior in noisy intermediate-scale quantum (NISQ) devices.

\section{Summary and Outlook} \label{sec:Outlook}

We have introduced Quantum Liouvillian Tomography (QLT), a novel protocol for reconstructing non-Markovian generators of open quantum dynamics. Unlike existing approaches that typically rely on assumptions of Markovianity or time-independence, QLT directly estimates time-local Liouvillians by regressing the time derivatives of experimentally measured Pauli string probabilities. 
In addition to these derivatives, the protocol relies on estimates of SPAM parameters and the associated CPTP map. This framework enables a fully data-driven characterization of dissipative processes and the identification of memory effects in quantum evolution.

Numerical benchmarks on synthetic data show that QLT accurately reconstructs Hamiltonians, dissipative rates, and the full Liouvillian spectrum. While the recovery of jump operators is more susceptible to noise, the method reliably captures the dominant dissipative channels and detects non-Markovian signatures, such as negative dissipation rates.

We also report the first experimental implementation of QLT using data from the Helmi quantum processor. Despite the challenges of limited control and inherent noise, the protocol successfully identified the residual Hamiltonian and revealed non-Markovian behavior in idle two-qubit circuits. These results highlight the potential of QLT to operate under realistic, NISQ conditions. 
Further improvements in hardware access—particularly pulse-level control—along with error mitigation strategies may significantly enhance the precision and applicability of the method.

Taken together, our findings establish QLT as a promising tool for the analysis of open quantum systems beyond the Markovian regime. 
We envision that QLT will become a key tool for probing complex quantum noise, informing the design of robust error correction strategies, and ultimately advancing our understanding of quantum devices operating in realistic environments.

\vspace{16pt}
DA and PR acknowledge support by FCT through Grant No. UID/CTM/04540/2020 to the I\&D unit Centro de Física e Engenharia de Materiais Avançados and by INCD funded by FCT and FEDER under the project 01/SAICT/2016  022153.  
This work was supported by Research Council of Norway, project “IKTPLUSS-IKT og digital innovasjon
- 333979” (KW and SD), and by FCT-Portugal, Grant Agreement No. 101017733 (DA and PR), as part of the QuantERA II project “DQUANT: A Dissipative Quantum Chaos perspective on Near-Term Quantum Computing”. \footnote{ https://doi.org/10.54499/QuantERA/0003/2021} 

\bibliography{apssamp}

\providecommand{\noopsort}[1]{}\providecommand{\singleletter}[1]{#1}%
\begin{thebibliography}{47}%
\makeatletter
\providecommand \@ifxundefined [1]{%
 \@ifx{#1\undefined}
}%
\providecommand \@ifnum [1]{%
 \ifnum #1\expandafter \@firstoftwo
 \else \expandafter \@secondoftwo
 \fi
}%
\providecommand \@ifx [1]{%
 \ifx #1\expandafter \@firstoftwo
 \else \expandafter \@secondoftwo
 \fi
}%
\providecommand \natexlab [1]{#1}%
\providecommand \enquote  [1]{``#1''}%
\providecommand \bibnamefont  [1]{#1}%
\providecommand \bibfnamefont [1]{#1}%
\providecommand \citenamefont [1]{#1}%
\providecommand \href@noop [0]{\@secondoftwo}%
\providecommand \href [0]{\begingroup \@sanitize@url \@href}%
\providecommand \@href[1]{\@@startlink{#1}\@@href}%
\providecommand \@@href[1]{\endgroup#1\@@endlink}%
\providecommand \@sanitize@url [0]{\catcode `\\12\catcode `\$12\catcode `\&12\catcode `\#12\catcode `\^12\catcode `\_12\catcode `\%12\relax}%
\providecommand \@@startlink[1]{}%
\providecommand \@@endlink[0]{}%
\providecommand \url  [0]{\begingroup\@sanitize@url \@url }%
\providecommand \@url [1]{\endgroup\@href {#1}{\urlprefix }}%
\providecommand \urlprefix  [0]{URL }%
\providecommand \Eprint [0]{\href }%
\providecommand \doibase [0]{https://doi.org/}%
\providecommand \selectlanguage [0]{\@gobble}%
\providecommand \bibinfo  [0]{\@secondoftwo}%
\providecommand \bibfield  [0]{\@secondoftwo}%
\providecommand \translation [1]{[#1]}%
\providecommand \BibitemOpen [0]{}%
\providecommand \bibitemStop [0]{}%
\providecommand \bibitemNoStop [0]{.\EOS\space}%
\providecommand \EOS [0]{\spacefactor3000\relax}%
\providecommand \BibitemShut  [1]{\csname bibitem#1\endcsname}%
\let\auto@bib@innerbib\@empty
\bibitem [{\citenamefont {Eisert}\ \emph {et~al.}(2020)\citenamefont {Eisert}, \citenamefont {Hangleiter}, \citenamefont {Walk}, \citenamefont {Roth}, \citenamefont {Markham}, \citenamefont {Parekh}, \citenamefont {Chabaud},\ and\ \citenamefont {Kashefi}}]{eisert_quantum_2020}%
  \BibitemOpen
  \bibfield  {author} {\bibinfo {author} {\bibfnamefont {J.}~\bibnamefont {Eisert}}, \bibinfo {author} {\bibfnamefont {D.}~\bibnamefont {Hangleiter}}, \bibinfo {author} {\bibfnamefont {N.}~\bibnamefont {Walk}}, \bibinfo {author} {\bibfnamefont {I.}~\bibnamefont {Roth}}, \bibinfo {author} {\bibfnamefont {D.}~\bibnamefont {Markham}}, \bibinfo {author} {\bibfnamefont {R.}~\bibnamefont {Parekh}}, \bibinfo {author} {\bibfnamefont {U.}~\bibnamefont {Chabaud}},\ and\ \bibinfo {author} {\bibfnamefont {E.}~\bibnamefont {Kashefi}},\ }\bibfield  {title} {\bibinfo {title} {Quantum certification and benchmarking},\ }\href {https://doi.org/10.1038/s42254-020-0186-4} {\bibfield  {journal} {\bibinfo  {journal} {Nature Reviews Physics}\ }\textbf {\bibinfo {volume} {2}},\ \bibinfo {pages} {382} (\bibinfo {year} {2020})}\BibitemShut {NoStop}%
\bibitem [{\citenamefont {Blume-Kohout}\ \emph {et~al.}(2025)\citenamefont {Blume-Kohout}, \citenamefont {Proctor},\ and\ \citenamefont {Young}}]{blume2025}%
  \BibitemOpen
  \bibfield  {author} {\bibinfo {author} {\bibfnamefont {R.}~\bibnamefont {Blume-Kohout}}, \bibinfo {author} {\bibfnamefont {T.}~\bibnamefont {Proctor}},\ and\ \bibinfo {author} {\bibfnamefont {K.}~\bibnamefont {Young}},\ }\bibfield  {title} {\bibinfo {title} {Quantum characterization, verification, and validation},\ }\href {https://arxiv.org/abs/2503.16383} {\bibfield  {journal} {\bibinfo  {journal} {arXiv preprint arXiv:2503.16383}\ } (\bibinfo {year} {2025})}\BibitemShut {NoStop}%
\bibitem [{\citenamefont {Gebhart}\ \emph {et~al.}(2023{\natexlab{a}})\citenamefont {Gebhart}, \citenamefont {Santagati}, \citenamefont {Gentile}, \citenamefont {Gauger}, \citenamefont {Craig}, \citenamefont {Ares}, \citenamefont {Banchi}, \citenamefont {Marquardt}, \citenamefont {Pezze'},\ and\ \citenamefont {Bonato}}]{gebhart_learning_2023}%
  \BibitemOpen
  \bibfield  {author} {\bibinfo {author} {\bibfnamefont {V.}~\bibnamefont {Gebhart}}, \bibinfo {author} {\bibfnamefont {R.}~\bibnamefont {Santagati}}, \bibinfo {author} {\bibfnamefont {A.~A.}\ \bibnamefont {Gentile}}, \bibinfo {author} {\bibfnamefont {E.~M.}\ \bibnamefont {Gauger}}, \bibinfo {author} {\bibfnamefont {D.}~\bibnamefont {Craig}}, \bibinfo {author} {\bibfnamefont {N.}~\bibnamefont {Ares}}, \bibinfo {author} {\bibfnamefont {L.}~\bibnamefont {Banchi}}, \bibinfo {author} {\bibfnamefont {F.}~\bibnamefont {Marquardt}}, \bibinfo {author} {\bibfnamefont {L.}~\bibnamefont {Pezze'}},\ and\ \bibinfo {author} {\bibfnamefont {C.}~\bibnamefont {Bonato}},\ }\bibfield  {title} {\bibinfo {title} {Learning {Quantum} {Systems}},\ }\href {https://doi.org/10.1038/s42254-022-00552-1} {\bibfield  {journal} {\bibinfo  {journal} {Nature Reviews Physics}\ }\textbf {\bibinfo {volume} {5}},\ \bibinfo {pages} {141} (\bibinfo {year} {2023}{\natexlab{a}})}\BibitemShut {NoStop}%
\bibitem [{\citenamefont {Magesan}\ \emph {et~al.}(2012)\citenamefont {Magesan}, \citenamefont {Gambetta},\ and\ \citenamefont {Emerson}}]{magesan_characterizing_2012}%
  \BibitemOpen
  \bibfield  {author} {\bibinfo {author} {\bibfnamefont {E.}~\bibnamefont {Magesan}}, \bibinfo {author} {\bibfnamefont {J.~M.}\ \bibnamefont {Gambetta}},\ and\ \bibinfo {author} {\bibfnamefont {J.}~\bibnamefont {Emerson}},\ }\bibfield  {title} {\bibinfo {title} {Characterizing quantum gates via randomized benchmarking},\ }\href {https://doi.org/10.1103/PhysRevA.85.042311} {\bibfield  {journal} {\bibinfo  {journal} {Physical Review A}\ }\textbf {\bibinfo {volume} {85}},\ \bibinfo {pages} {042311} (\bibinfo {year} {2012})}\BibitemShut {NoStop}%
\bibitem [{\citenamefont {Mohseni}\ \emph {et~al.}(2008)\citenamefont {Mohseni}, \citenamefont {Rezakhani},\ and\ \citenamefont {Lidar}}]{mohseni_quantum_2008}%
  \BibitemOpen
  \bibfield  {author} {\bibinfo {author} {\bibfnamefont {M.}~\bibnamefont {Mohseni}}, \bibinfo {author} {\bibfnamefont {A.~T.}\ \bibnamefont {Rezakhani}},\ and\ \bibinfo {author} {\bibfnamefont {D.~A.}\ \bibnamefont {Lidar}},\ }\bibfield  {title} {\bibinfo {title} {Quantum {Process} {Tomography}: {Resource} {Analysis} of {Different} {Strategies}},\ }\href {https://doi.org/10.1103/PhysRevA.77.032322} {\bibfield  {journal} {\bibinfo  {journal} {Physical Review A}\ }\textbf {\bibinfo {volume} {77}},\ \bibinfo {pages} {032322} (\bibinfo {year} {2008})}\BibitemShut {NoStop}%
\bibitem [{\citenamefont {Kiktenko}\ \emph {et~al.}(2020)\citenamefont {Kiktenko}, \citenamefont {Kublikova},\ and\ \citenamefont {Fedorov}}]{kiktenko_estimating_2020}%
  \BibitemOpen
  \bibfield  {author} {\bibinfo {author} {\bibfnamefont {E.~O.}\ \bibnamefont {Kiktenko}}, \bibinfo {author} {\bibfnamefont {D.~N.}\ \bibnamefont {Kublikova}},\ and\ \bibinfo {author} {\bibfnamefont {A.~K.}\ \bibnamefont {Fedorov}},\ }\bibfield  {title} {\bibinfo {title} {Estimating the precision for quantum process tomography},\ }\href {https://doi.org/10.1117/1.OE.59.6.061614} {\bibfield  {journal} {\bibinfo  {journal} {Optical Engineering}\ }\textbf {\bibinfo {volume} {59}},\ \bibinfo {pages} {1} (\bibinfo {year} {2020})}\BibitemShut {NoStop}%
\bibitem [{\citenamefont {Watrous}(2018{\natexlab{a}})}]{watrous2018}%
  \BibitemOpen
  \bibfield  {author} {\bibinfo {author} {\bibfnamefont {J.}~\bibnamefont {Watrous}},\ }\href {https://doi.org/10.1017/9781316848142} {\emph {\bibinfo {title} {The Theory of Quantum Information}}}\ (\bibinfo  {publisher} {Cambridge University Press},\ \bibinfo {address} {Cambridge},\ \bibinfo {year} {2018})\BibitemShut {NoStop}%
\bibitem [{\citenamefont {Holevo}(2019)}]{holevo2019}%
  \BibitemOpen
  \bibfield  {author} {\bibinfo {author} {\bibfnamefont {A.~S.}\ \bibnamefont {Holevo}},\ }\href {https://doi.org/10.1515/9783110642375} {\emph {\bibinfo {title} {Quantum Systems, Channels, Information}}}\ (\bibinfo  {publisher} {De Gruyter},\ \bibinfo {address} {Berlin},\ \bibinfo {year} {2019})\BibitemShut {NoStop}%
\bibitem [{\citenamefont {Preskill}(2018)}]{preskill_quantum_2018}%
  \BibitemOpen
  \bibfield  {author} {\bibinfo {author} {\bibfnamefont {J.}~\bibnamefont {Preskill}},\ }\bibfield  {title} {\bibinfo {title} {Quantum {Computing} in the {NISQ} era and beyond},\ }\href {https://doi.org/10.22331/q-2018-08-06-79} {\bibfield  {journal} {\bibinfo  {journal} {Quantum}\ }\textbf {\bibinfo {volume} {2}},\ \bibinfo {pages} {79} (\bibinfo {year} {2018})}\BibitemShut {NoStop}%
\bibitem [{\citenamefont {Foucart}\ and\ \citenamefont {Rauhut}(2013)}]{Foucart2013}%
  \BibitemOpen
  \bibfield  {author} {\bibinfo {author} {\bibfnamefont {S.}~\bibnamefont {Foucart}}\ and\ \bibinfo {author} {\bibfnamefont {H.}~\bibnamefont {Rauhut}},\ }\href {https://doi.org/10.1007/978-0-8176-4948-7} {\emph {\bibinfo {title} {A Mathematical Introduction to Compressive Sensing}}},\ Applied and Numerical Harmonic Analysis\ (\bibinfo  {publisher} {Birkhäuser},\ \bibinfo {year} {2013})\BibitemShut {NoStop}%
\bibitem [{\citenamefont {Gebhart}\ \emph {et~al.}(2023{\natexlab{b}})\citenamefont {Gebhart}, \citenamefont {Santagati}, \citenamefont {Gentile}, \citenamefont {Gauger}, \citenamefont {Craig}, \citenamefont {Ares}, \citenamefont {Banchi}, \citenamefont {Marquardt}, \citenamefont {Pezzè},\ and\ \citenamefont {Bonato}}]{gebhart2023}%
  \BibitemOpen
  \bibfield  {author} {\bibinfo {author} {\bibfnamefont {V.}~\bibnamefont {Gebhart}}, \bibinfo {author} {\bibfnamefont {R.}~\bibnamefont {Santagati}}, \bibinfo {author} {\bibfnamefont {A.~A.}\ \bibnamefont {Gentile}}, \bibinfo {author} {\bibfnamefont {E.~M.}\ \bibnamefont {Gauger}}, \bibinfo {author} {\bibfnamefont {D.}~\bibnamefont {Craig}}, \bibinfo {author} {\bibfnamefont {N.}~\bibnamefont {Ares}}, \bibinfo {author} {\bibfnamefont {L.}~\bibnamefont {Banchi}}, \bibinfo {author} {\bibfnamefont {F.}~\bibnamefont {Marquardt}}, \bibinfo {author} {\bibfnamefont {L.}~\bibnamefont {Pezzè}},\ and\ \bibinfo {author} {\bibfnamefont {C.}~\bibnamefont {Bonato}},\ }\bibfield  {title} {\bibinfo {title} {Learning quantum systems},\ }\href {https://doi.org/10.1038/s42254-022-00552-1} {\bibfield  {journal} {\bibinfo  {journal} {Nature Reviews Physics}\ }\textbf {\bibinfo {volume} {5}},\ \bibinfo {pages} {141} (\bibinfo {year} {2023}{\natexlab{b}})}\BibitemShut {NoStop}%
\bibitem [{\citenamefont {Torlai}\ \emph {et~al.}(2020)\citenamefont {Torlai}, \citenamefont {Carrasquilla}, \citenamefont {Aolita},\ and\ \citenamefont {Melko}}]{Torlai2020}%
  \BibitemOpen
  \bibfield  {author} {\bibinfo {author} {\bibfnamefont {G.}~\bibnamefont {Torlai}}, \bibinfo {author} {\bibfnamefont {J.}~\bibnamefont {Carrasquilla}}, \bibinfo {author} {\bibfnamefont {L.}~\bibnamefont {Aolita}},\ and\ \bibinfo {author} {\bibfnamefont {R.~G.}\ \bibnamefont {Melko}},\ }\bibfield  {title} {\bibinfo {title} {Modeling quantum processes with tensor networks: scalable quantum process tomography},\ }\href {https://doi.org/10.1038/s41534-020-0257-3} {\bibfield  {journal} {\bibinfo  {journal} {npj Quantum Information}\ }\textbf {\bibinfo {volume} {6}},\ \bibinfo {pages} {19} (\bibinfo {year} {2020})}\BibitemShut {NoStop}%
\bibitem [{\citenamefont {et~al.}(2011)}]{Shabani2011}%
  \BibitemOpen
  \bibfield  {author} {\bibinfo {author} {\bibfnamefont {A.~S.}\ \bibnamefont {et~al.}},\ }\bibfield  {title} {\bibinfo {title} {Efficient measurement of quantum dynamics via compressive sensing},\ }\href {https://doi.org/10.1103/PhysRevLett.106.100401} {\bibfield  {journal} {\bibinfo  {journal} {Phys. Rev. Lett.}\ }\textbf {\bibinfo {volume} {106}},\ \bibinfo {pages} {100401} (\bibinfo {year} {2011})}\BibitemShut {NoStop}%
\bibitem [{\citenamefont {et~al.}(2023)}]{Torlai2023}%
  \BibitemOpen
  \bibfield  {author} {\bibinfo {author} {\bibfnamefont {G.~T.}\ \bibnamefont {et~al.}},\ }\bibfield  {title} {\bibinfo {title} {Quantum process tomography with unsupervised learning and tensor networks},\ }\href {https://doi.org/10.1038/s41467-023-38332-9} {\bibfield  {journal} {\bibinfo  {journal} {Nat. Commun.}\ }\textbf {\bibinfo {volume} {14}},\ \bibinfo {pages} {2858} (\bibinfo {year} {2023})}\BibitemShut {NoStop}%
\bibitem [{\citenamefont {et~al.}(2024{\natexlab{a}})}]{DiColandrea2024}%
  \BibitemOpen
  \bibfield  {author} {\bibinfo {author} {\bibfnamefont {F.~D.~C.}\ \bibnamefont {et~al.}},\ }\bibfield  {title} {\bibinfo {title} {Fourier quantum process tomography},\ }\href {https://doi.org/10.1038/s41534-024-00844-7} {\bibfield  {journal} {\bibinfo  {journal} {Npj Quantum Inf.}\ }\textbf {\bibinfo {volume} {10}},\ \bibinfo {pages} {49} (\bibinfo {year} {2024}{\natexlab{a}})}\BibitemShut {NoStop}%
\bibitem [{\citenamefont {R.~Levy}\ and\ \citenamefont {Clark}(2024)}]{Levy2024}%
  \BibitemOpen
  \bibfield  {author} {\bibinfo {author} {\bibfnamefont {D.~L.}\ \bibnamefont {R.~Levy}}\ and\ \bibinfo {author} {\bibfnamefont {B.~K.}\ \bibnamefont {Clark}},\ }\bibfield  {title} {\bibinfo {title} {Classical shadows for quantum process tomography on near-term quantum computers},\ }\href {https://doi.org/10.1103/PhysRevResearch.6.013029} {\bibfield  {journal} {\bibinfo  {journal} {Phys. Rev. Res.}\ }\textbf {\bibinfo {volume} {6}},\ \bibinfo {pages} {013029} (\bibinfo {year} {2024})}\BibitemShut {NoStop}%
\bibitem [{\citenamefont {Onorati}\ \emph {et~al.}(2023)\citenamefont {Onorati}, \citenamefont {Kohler},\ and\ \citenamefont {Cubitt}}]{onorati_fitting_2023}%
  \BibitemOpen
  \bibfield  {author} {\bibinfo {author} {\bibfnamefont {E.}~\bibnamefont {Onorati}}, \bibinfo {author} {\bibfnamefont {T.}~\bibnamefont {Kohler}},\ and\ \bibinfo {author} {\bibfnamefont {T.~S.}\ \bibnamefont {Cubitt}},\ }\bibfield  {title} {\bibinfo {title} {Fitting quantum noise models to tomography data},\ }\href {https://doi.org/10.22331/q-2023-12-05-1197} {\bibfield  {journal} {\bibinfo  {journal} {Quantum}\ }\textbf {\bibinfo {volume} {7}},\ \bibinfo {pages} {1197} (\bibinfo {year} {2023})}\BibitemShut {NoStop}%
\bibitem [{\citenamefont {Samach}\ \emph {et~al.}(2022)\citenamefont {Samach}, \citenamefont {Greene}, \citenamefont {Borregaard}, \citenamefont {Christandl}, \citenamefont {Barreto}, \citenamefont {Kim}, \citenamefont {McNally}, \citenamefont {Melville}, \citenamefont {Niedzielski}, \citenamefont {Sung}, \citenamefont {Rosenberg}, \citenamefont {Schwartz}, \citenamefont {Yoder}, \citenamefont {Orlando}, \citenamefont {Wang}, \citenamefont {Gustavsson}, \citenamefont {Kjaergaard},\ and\ \citenamefont {Oliver}}]{samach_lindblad_2022}%
  \BibitemOpen
  \bibfield  {author} {\bibinfo {author} {\bibfnamefont {G.~O.}\ \bibnamefont {Samach}}, \bibinfo {author} {\bibfnamefont {A.}~\bibnamefont {Greene}}, \bibinfo {author} {\bibfnamefont {J.}~\bibnamefont {Borregaard}}, \bibinfo {author} {\bibfnamefont {M.}~\bibnamefont {Christandl}}, \bibinfo {author} {\bibfnamefont {J.}~\bibnamefont {Barreto}}, \bibinfo {author} {\bibfnamefont {D.~K.}\ \bibnamefont {Kim}}, \bibinfo {author} {\bibfnamefont {C.~M.}\ \bibnamefont {McNally}}, \bibinfo {author} {\bibfnamefont {A.}~\bibnamefont {Melville}}, \bibinfo {author} {\bibfnamefont {B.~M.}\ \bibnamefont {Niedzielski}}, \bibinfo {author} {\bibfnamefont {Y.}~\bibnamefont {Sung}}, \bibinfo {author} {\bibfnamefont {D.}~\bibnamefont {Rosenberg}}, \bibinfo {author} {\bibfnamefont {M.~E.}\ \bibnamefont {Schwartz}}, \bibinfo {author} {\bibfnamefont {J.~L.}\ \bibnamefont {Yoder}}, \bibinfo {author} {\bibfnamefont {T.~P.}\ \bibnamefont {Orlando}}, \bibinfo {author} {\bibfnamefont {J.~I.-J.}\ \bibnamefont {Wang}}, \bibinfo {author}
  {\bibfnamefont {S.}~\bibnamefont {Gustavsson}}, \bibinfo {author} {\bibfnamefont {M.}~\bibnamefont {Kjaergaard}},\ and\ \bibinfo {author} {\bibfnamefont {W.~D.}\ \bibnamefont {Oliver}},\ }\bibfield  {title} {\bibinfo {title} {Lindblad {Tomography} of a {Superconducting} {Quantum} {Processor}},\ }\href {https://doi.org/10.1103/PhysRevApplied.18.064056} {\bibfield  {journal} {\bibinfo  {journal} {Physical Review Applied}\ }\textbf {\bibinfo {volume} {18}},\ \bibinfo {pages} {064056} (\bibinfo {year} {2022})}\BibitemShut {NoStop}%
\bibitem [{\citenamefont {Olsacher}\ \emph {et~al.}(2024)\citenamefont {Olsacher}, \citenamefont {Kraft}, \citenamefont {Kokail}, \citenamefont {Kraus},\ and\ \citenamefont {Zoller}}]{olsacher_hamiltonian_2024}%
  \BibitemOpen
  \bibfield  {author} {\bibinfo {author} {\bibfnamefont {T.}~\bibnamefont {Olsacher}}, \bibinfo {author} {\bibfnamefont {T.}~\bibnamefont {Kraft}}, \bibinfo {author} {\bibfnamefont {C.}~\bibnamefont {Kokail}}, \bibinfo {author} {\bibfnamefont {B.}~\bibnamefont {Kraus}},\ and\ \bibinfo {author} {\bibfnamefont {P.}~\bibnamefont {Zoller}},\ }\href {http://arxiv.org/abs/2405.06768} {\bibinfo {title} {Hamiltonian and {Liouvillian} learning in weakly-dissipative quantum many-body systems}} (\bibinfo {year} {2024})\BibitemShut {NoStop}%
\bibitem [{\citenamefont {Howard}\ \emph {et~al.}(2006)\citenamefont {Howard}, \citenamefont {Twamley}, \citenamefont {Wittmann}, \citenamefont {Gaebel}, \citenamefont {Jelezko},\ and\ \citenamefont {Wrachtrup}}]{howard_quantum_2006}%
  \BibitemOpen
  \bibfield  {author} {\bibinfo {author} {\bibfnamefont {M.}~\bibnamefont {Howard}}, \bibinfo {author} {\bibfnamefont {J.}~\bibnamefont {Twamley}}, \bibinfo {author} {\bibfnamefont {C.}~\bibnamefont {Wittmann}}, \bibinfo {author} {\bibfnamefont {T.}~\bibnamefont {Gaebel}}, \bibinfo {author} {\bibfnamefont {F.}~\bibnamefont {Jelezko}},\ and\ \bibinfo {author} {\bibfnamefont {J.}~\bibnamefont {Wrachtrup}},\ }\bibfield  {title} {\bibinfo {title} {Quantum process tomography and {Linblad} estimation of a solid state qubit},\ }\href {https://doi.org/10.1088/1367-2630/8/3/033} {\bibfield  {journal} {\bibinfo  {journal} {New Journal of Physics}\ }\textbf {\bibinfo {volume} {8}},\ \bibinfo {pages} {33} (\bibinfo {year} {2006})}\BibitemShut {NoStop}%
\bibitem [{\citenamefont {Pastori}\ \emph {et~al.}(2022)\citenamefont {Pastori}, \citenamefont {Olsacher}, \citenamefont {Kokail},\ and\ \citenamefont {Zoller}}]{pastori_characterization_2022}%
  \BibitemOpen
  \bibfield  {author} {\bibinfo {author} {\bibfnamefont {L.}~\bibnamefont {Pastori}}, \bibinfo {author} {\bibfnamefont {T.}~\bibnamefont {Olsacher}}, \bibinfo {author} {\bibfnamefont {C.}~\bibnamefont {Kokail}},\ and\ \bibinfo {author} {\bibfnamefont {P.}~\bibnamefont {Zoller}},\ }\bibfield  {title} {\bibinfo {title} {Characterization and {Verification} of {Trotterized} {Digital} {Quantum} {Simulation} via {Hamiltonian} and {Liouvillian} {Learning}},\ }\href {https://doi.org/10.1103/PRXQuantum.3.030324} {\bibfield  {journal} {\bibinfo  {journal} {PRX Quantum}\ }\textbf {\bibinfo {volume} {3}},\ \bibinfo {pages} {030324} (\bibinfo {year} {2022})}\BibitemShut {NoStop}%
\bibitem [{\citenamefont {Gorini}\ \emph {et~al.}(1976)\citenamefont {Gorini}, \citenamefont {Kossakowski},\ and\ \citenamefont {Sudarshan}}]{GKS}%
  \BibitemOpen
  \bibfield  {author} {\bibinfo {author} {\bibfnamefont {V.}~\bibnamefont {Gorini}}, \bibinfo {author} {\bibfnamefont {A.}~\bibnamefont {Kossakowski}},\ and\ \bibinfo {author} {\bibfnamefont {E.~C.~G.}\ \bibnamefont {Sudarshan}},\ }\bibfield  {title} {\bibinfo {title} {Completely positive dynamical semigroups of n-level systems},\ }\href {https://doi.org/10.1063/1.522979} {\bibfield  {journal} {\bibinfo  {journal} {Journal of Mathematical Physics}\ }\textbf {\bibinfo {volume} {17}},\ \bibinfo {pages} {821} (\bibinfo {year} {1976})}\BibitemShut {NoStop}%
\bibitem [{\citenamefont {Lindblad}(1976)}]{L}%
  \BibitemOpen
  \bibfield  {author} {\bibinfo {author} {\bibfnamefont {G.}~\bibnamefont {Lindblad}},\ }\bibfield  {title} {\bibinfo {title} {On the generators of quantum dynamical semigroups},\ }\href {https://doi.org/10.1007/BF01608499} {\bibfield  {journal} {\bibinfo  {journal} {Communications in Mathematical Physics}\ }\textbf {\bibinfo {volume} {48}},\ \bibinfo {pages} {119} (\bibinfo {year} {1976})}\BibitemShut {NoStop}%
\bibitem [{\citenamefont {et~al.}(2024{\natexlab{b}})}]{StilckFranca2024}%
  \BibitemOpen
  \bibfield  {author} {\bibinfo {author} {\bibfnamefont {D.~S.~F.}\ \bibnamefont {et~al.}},\ }\bibfield  {title} {\bibinfo {title} {Efficient and robust estimation of many-qubit hamiltonians},\ }\href {https://doi.org/10.1038/s41467-023-44012-5} {\bibfield  {journal} {\bibinfo  {journal} {Nat. Commun.}\ }\textbf {\bibinfo {volume} {15}},\ \bibinfo {pages} {311} (\bibinfo {year} {2024}{\natexlab{b}})}\BibitemShut {NoStop}%
\bibitem [{\citenamefont {Ángel Rivas}\ \emph {et~al.}(2014)\citenamefont {Ángel Rivas}, \citenamefont {Huelga},\ and\ \citenamefont {Plenio}}]{rivas2014}%
  \BibitemOpen
  \bibfield  {author} {\bibinfo {author} {\bibnamefont {Ángel Rivas}}, \bibinfo {author} {\bibfnamefont {S.~F.}\ \bibnamefont {Huelga}},\ and\ \bibinfo {author} {\bibfnamefont {M.~B.}\ \bibnamefont {Plenio}},\ }\bibfield  {title} {\bibinfo {title} {Quantum non-markovianity: Characterization, quantification and detection},\ }\href {https://doi.org/10.1088/0034-4885/77/9/094001} {\bibfield  {journal} {\bibinfo  {journal} {Reports on Progress in Physics}\ }\textbf {\bibinfo {volume} {77}},\ \bibinfo {pages} {094001} (\bibinfo {year} {2014})}\BibitemShut {NoStop}%
\bibitem [{\citenamefont {Breuer}\ \emph {et~al.}(2016{\natexlab{a}})\citenamefont {Breuer}, \citenamefont {Laine}, \citenamefont {Piilo},\ and\ \citenamefont {Vacchini}}]{Breuer2016}%
  \BibitemOpen
  \bibfield  {author} {\bibinfo {author} {\bibfnamefont {H.-P.}\ \bibnamefont {Breuer}}, \bibinfo {author} {\bibfnamefont {E.-M.}\ \bibnamefont {Laine}}, \bibinfo {author} {\bibfnamefont {J.}~\bibnamefont {Piilo}},\ and\ \bibinfo {author} {\bibfnamefont {B.}~\bibnamefont {Vacchini}},\ }\bibfield  {title} {\bibinfo {title} {Colloquium: Non-markovian dynamics in open quantum systems},\ }\href {https://doi.org/10.1103/RevModPhys.88.021002} {\bibfield  {journal} {\bibinfo  {journal} {Reviews of Modern Physics}\ }\textbf {\bibinfo {volume} {88}},\ \bibinfo {pages} {021002} (\bibinfo {year} {2016}{\natexlab{a}})}\BibitemShut {NoStop}%
\bibitem [{\citenamefont {Chru{\'s}ci{\'n}ski}(2022)}]{Chruscinski2022}%
  \BibitemOpen
  \bibfield  {author} {\bibinfo {author} {\bibfnamefont {D.}~\bibnamefont {Chru{\'s}ci{\'n}ski}},\ }\bibfield  {title} {\bibinfo {title} {Dynamical maps beyond markovian regime},\ }\href {https://doi.org/10.1016/j.physrep.2022.09.002} {\bibfield  {journal} {\bibinfo  {journal} {Physics Reports}\ }\textbf {\bibinfo {volume} {992}},\ \bibinfo {pages} {1} (\bibinfo {year} {2022})}\BibitemShut {NoStop}%
\bibitem [{\citenamefont {White}\ \emph {et~al.}(2020)\citenamefont {White}, \citenamefont {Hill}, \citenamefont {Pollock}, \citenamefont {Hollenberg},\ and\ \citenamefont {Modi}}]{white2020}%
  \BibitemOpen
  \bibfield  {author} {\bibinfo {author} {\bibfnamefont {G.~A.~L.}\ \bibnamefont {White}}, \bibinfo {author} {\bibfnamefont {C.~D.}\ \bibnamefont {Hill}}, \bibinfo {author} {\bibfnamefont {F.~A.}\ \bibnamefont {Pollock}}, \bibinfo {author} {\bibfnamefont {L.~C.~L.}\ \bibnamefont {Hollenberg}},\ and\ \bibinfo {author} {\bibfnamefont {K.}~\bibnamefont {Modi}},\ }\bibfield  {title} {\bibinfo {title} {Demonstration of non-markovian process characterisation and control on a quantum processor},\ }\href {https://doi.org/10.1038/s41467-020-20113-3} {\bibfield  {journal} {\bibinfo  {journal} {Nature Communications}\ }\textbf {\bibinfo {volume} {11}},\ \bibinfo {pages} {6301} (\bibinfo {year} {2020})}\BibitemShut {NoStop}%
\bibitem [{\citenamefont {White}\ \emph {et~al.}(2022)\citenamefont {White}, \citenamefont {Pollock}, \citenamefont {Hollenberg}, \citenamefont {Modi},\ and\ \citenamefont {Hill}}]{white_non-markovian_2022}%
  \BibitemOpen
  \bibfield  {author} {\bibinfo {author} {\bibfnamefont {G.}~\bibnamefont {White}}, \bibinfo {author} {\bibfnamefont {F.}~\bibnamefont {Pollock}}, \bibinfo {author} {\bibfnamefont {L.}~\bibnamefont {Hollenberg}}, \bibinfo {author} {\bibfnamefont {K.}~\bibnamefont {Modi}},\ and\ \bibinfo {author} {\bibfnamefont {C.}~\bibnamefont {Hill}},\ }\bibfield  {title} {\bibinfo {title} {Non-{Markovian} {Quantum} {Process} {Tomography}},\ }\href {https://doi.org/10.1103/PRXQuantum.3.020344} {\bibfield  {journal} {\bibinfo  {journal} {PRX Quantum}\ }\textbf {\bibinfo {volume} {3}},\ \bibinfo {pages} {020344} (\bibinfo {year} {2022})}\BibitemShut {NoStop}%
\bibitem [{\citenamefont {Pollock}\ \emph {et~al.}(2018)\citenamefont {Pollock}, \citenamefont {Rodríguez-Rosario}, \citenamefont {Frauenheim}, \citenamefont {Paternostro},\ and\ \citenamefont {Modi}}]{pollock2018nonmarkovian}%
  \BibitemOpen
  \bibfield  {author} {\bibinfo {author} {\bibfnamefont {F.~A.}\ \bibnamefont {Pollock}}, \bibinfo {author} {\bibfnamefont {C.}~\bibnamefont {Rodríguez-Rosario}}, \bibinfo {author} {\bibfnamefont {T.}~\bibnamefont {Frauenheim}}, \bibinfo {author} {\bibfnamefont {M.}~\bibnamefont {Paternostro}},\ and\ \bibinfo {author} {\bibfnamefont {K.}~\bibnamefont {Modi}},\ }\bibfield  {title} {\bibinfo {title} {Non-markovian quantum processes: Complete framework and efficient characterization},\ }\href {https://doi.org/10.1103/PhysRevA.97.012127} {\bibfield  {journal} {\bibinfo  {journal} {Physical Review A}\ }\textbf {\bibinfo {volume} {97}},\ \bibinfo {pages} {012127} (\bibinfo {year} {2018})}\BibitemShut {NoStop}%
\bibitem [{\citenamefont {{VTT Technical Research Centre of Finland}}(2024)}]{helmiQC}%
  \BibitemOpen
  \bibfield  {author} {\bibinfo {author} {\bibnamefont {{VTT Technical Research Centre of Finland}}},\ }\href@noop {} {\bibinfo {title} {{Helmi: A superconducting 5-qubit quantum computer}}},\ \bibinfo {howpublished} {\url{https://vttresearch.github.io/quantum-computer-documentation/helmi/}} (\bibinfo {year} {2024})\BibitemShut {NoStop}%
\bibitem [{\citenamefont {Hall}\ \emph {et~al.}(2014)\citenamefont {Hall}, \citenamefont {Cresser}, \citenamefont {Li},\ and\ \citenamefont {Andersson}}]{hall_canonical_2014}%
  \BibitemOpen
  \bibfield  {author} {\bibinfo {author} {\bibfnamefont {M.~J.~W.}\ \bibnamefont {Hall}}, \bibinfo {author} {\bibfnamefont {J.~D.}\ \bibnamefont {Cresser}}, \bibinfo {author} {\bibfnamefont {L.}~\bibnamefont {Li}},\ and\ \bibinfo {author} {\bibfnamefont {E.}~\bibnamefont {Andersson}},\ }\bibfield  {title} {\bibinfo {title} {Canonical form of master equations and characterization of non-{Markovianity}},\ }\href {https://doi.org/10.1103/PhysRevA.89.042120} {\bibfield  {journal} {\bibinfo  {journal} {Physical Review A}\ }\textbf {\bibinfo {volume} {89}},\ \bibinfo {pages} {042120} (\bibinfo {year} {2014})}\BibitemShut {NoStop}%
\bibitem [{\citenamefont {Manzano}(2020)}]{manzano_short_2020}%
  \BibitemOpen
  \bibfield  {author} {\bibinfo {author} {\bibfnamefont {D.}~\bibnamefont {Manzano}},\ }\bibfield  {title} {\bibinfo {title} {A short introduction to the {Lindblad} master equation},\ }\href {https://doi.org/10.1063/1.5115323} {\bibfield  {journal} {\bibinfo  {journal} {AIP Advances}\ }\textbf {\bibinfo {volume} {10}},\ \bibinfo {pages} {025106} (\bibinfo {year} {2020})}\BibitemShut {NoStop}%
\bibitem [{\citenamefont {Breuer}\ \emph {et~al.}(2016{\natexlab{b}})\citenamefont {Breuer}, \citenamefont {Laine}, \citenamefont {Piilo},\ and\ \citenamefont {Vacchini}}]{breuer_non-markovian_2016}%
  \BibitemOpen
  \bibfield  {author} {\bibinfo {author} {\bibfnamefont {H.-P.}\ \bibnamefont {Breuer}}, \bibinfo {author} {\bibfnamefont {E.-M.}\ \bibnamefont {Laine}}, \bibinfo {author} {\bibfnamefont {J.}~\bibnamefont {Piilo}},\ and\ \bibinfo {author} {\bibfnamefont {B.}~\bibnamefont {Vacchini}},\ }\bibfield  {title} {\bibinfo {title} {Non-{Markovian} dynamics in open quantum systems},\ }\href {https://doi.org/10.1103/RevModPhys.88.021002} {\bibfield  {journal} {\bibinfo  {journal} {Reviews of Modern Physics}\ }\textbf {\bibinfo {volume} {88}},\ \bibinfo {pages} {021002} (\bibinfo {year} {2016}{\natexlab{b}})}\BibitemShut {NoStop}%
\bibitem [{\citenamefont {Kay}(2023)}]{kay2023tutorialquantikzpackage}%
  \BibitemOpen
  \bibfield  {author} {\bibinfo {author} {\bibfnamefont {A.}~\bibnamefont {Kay}},\ }\href {https://arxiv.org/abs/1809.03842} {\bibinfo {title} {Tutorial on the quantikz package}} (\bibinfo {year} {2023}),\ \Eprint {https://arxiv.org/abs/1809.03842} {arXiv:1809.03842 [quant-ph]} \BibitemShut {NoStop}%
\bibitem [{\citenamefont {Wold}\ \emph {et~al.}(2024)\citenamefont {Wold}, \citenamefont {Ribeiro},\ and\ \citenamefont {Denisov}}]{wold_universal_2024}%
  \BibitemOpen
  \bibfield  {author} {\bibinfo {author} {\bibfnamefont {K.}~\bibnamefont {Wold}}, \bibinfo {author} {\bibfnamefont {P.}~\bibnamefont {Ribeiro}},\ and\ \bibinfo {author} {\bibfnamefont {S.}~\bibnamefont {Denisov}},\ }\href {http://arxiv.org/abs/2405.11625} {\bibinfo {title} {Universal spectra of noisy parameterized quantum circuits}} (\bibinfo {year} {2024})\BibitemShut {NoStop}%
\bibitem [{\citenamefont {Roncallo}\ \emph {et~al.}(2024)\citenamefont {Roncallo}, \citenamefont {Maccone},\ and\ \citenamefont {Macchiavello}}]{roncallo2024pauli}%
  \BibitemOpen
  \bibfield  {author} {\bibinfo {author} {\bibfnamefont {S.}~\bibnamefont {Roncallo}}, \bibinfo {author} {\bibfnamefont {L.}~\bibnamefont {Maccone}},\ and\ \bibinfo {author} {\bibfnamefont {C.}~\bibnamefont {Macchiavello}},\ }\bibfield  {title} {\bibinfo {title} {Pauli transfer matrix direct reconstruction: Channel characterization without full process tomography},\ }\href {https://doi.org/10.1088/2058-9565/ad04e7} {\bibfield  {journal} {\bibinfo  {journal} {Quantum Science and Technology}\ }\textbf {\bibinfo {volume} {9}},\ \bibinfo {pages} {015010} (\bibinfo {year} {2024})}\BibitemShut {NoStop}%
\bibitem [{\citenamefont {van~den Berg}\ \emph {et~al.}(2023)\citenamefont {van~den Berg}, \citenamefont {Minev}, \citenamefont {Kandala},\ and\ \citenamefont {Temme}}]{vanDenBerg2023probabilistic}%
  \BibitemOpen
  \bibfield  {author} {\bibinfo {author} {\bibfnamefont {E.}~\bibnamefont {van~den Berg}}, \bibinfo {author} {\bibfnamefont {Z.~K.}\ \bibnamefont {Minev}}, \bibinfo {author} {\bibfnamefont {A.}~\bibnamefont {Kandala}},\ and\ \bibinfo {author} {\bibfnamefont {K.}~\bibnamefont {Temme}},\ }\bibfield  {title} {\bibinfo {title} {Probabilistic error cancellation with sparse pauli–lindblad models on noisy quantum processors},\ }\href {https://doi.org/10.1038/s41567-023-02042-2} {\bibfield  {journal} {\bibinfo  {journal} {Nature Physics}\ }\textbf {\bibinfo {volume} {19}},\ \bibinfo {pages} {1116} (\bibinfo {year} {2023})}\BibitemShut {NoStop}%
\bibitem [{\citenamefont {Kingma}\ and\ \citenamefont {Ba}(2017)}]{kingma2017adammethodstochasticoptimization}%
  \BibitemOpen
  \bibfield  {author} {\bibinfo {author} {\bibfnamefont {D.~P.}\ \bibnamefont {Kingma}}\ and\ \bibinfo {author} {\bibfnamefont {J.}~\bibnamefont {Ba}},\ }\href {https://arxiv.org/abs/1412.6980} {\bibinfo {title} {Adam: A method for stochastic optimization}} (\bibinfo {year} {2017}),\ \Eprint {https://arxiv.org/abs/1412.6980} {arXiv:1412.6980 [cs.LG]} \BibitemShut {NoStop}%
\bibitem [{\citenamefont {Chruscinski}\ and\ \citenamefont {Kossakowski}(2010)}]{chruscinski_non-markovian_2010}%
  \BibitemOpen
  \bibfield  {author} {\bibinfo {author} {\bibfnamefont {D.}~\bibnamefont {Chruscinski}}\ and\ \bibinfo {author} {\bibfnamefont {A.}~\bibnamefont {Kossakowski}},\ }\bibfield  {title} {\bibinfo {title} {Non-{Markovian} quantum dynamics: local versus non-local},\ }\href {https://doi.org/10.1103/PhysRevLett.104.070406} {\bibfield  {journal} {\bibinfo  {journal} {Physical Review Letters}\ }\textbf {\bibinfo {volume} {104}},\ \bibinfo {pages} {070406} (\bibinfo {year} {2010})}\BibitemShut {NoStop}%
\bibitem [{\citenamefont {Watrous}(2018{\natexlab{b}})}]{watrous_theory_2018}%
  \BibitemOpen
  \bibfield  {author} {\bibinfo {author} {\bibfnamefont {J.}~\bibnamefont {Watrous}},\ }\href {https://doi.org/10.1017/9781316848142} {\emph {\bibinfo {title} {The {Theory} of {Quantum} {Information}}}}\ (\bibinfo  {publisher} {Cambridge University Press},\ \bibinfo {year} {2018})\BibitemShut {NoStop}%
\bibitem [{\citenamefont {McKay}\ \emph {et~al.}(2017)\citenamefont {McKay}, \citenamefont {Wood}, \citenamefont {Sheldon}, \citenamefont {Chow},\ and\ \citenamefont {Gambetta}}]{mckay_efficient_2017}%
  \BibitemOpen
  \bibfield  {author} {\bibinfo {author} {\bibfnamefont {D.~C.}\ \bibnamefont {McKay}}, \bibinfo {author} {\bibfnamefont {C.~J.}\ \bibnamefont {Wood}}, \bibinfo {author} {\bibfnamefont {S.}~\bibnamefont {Sheldon}}, \bibinfo {author} {\bibfnamefont {J.~M.}\ \bibnamefont {Chow}},\ and\ \bibinfo {author} {\bibfnamefont {J.~M.}\ \bibnamefont {Gambetta}},\ }\bibfield  {title} {\bibinfo {title} {Efficient {Z} gates for quantum computing},\ }\href {https://doi.org/10.1103/PhysRevA.96.022330} {\bibfield  {journal} {\bibinfo  {journal} {Physical Review A}\ }\textbf {\bibinfo {volume} {96}},\ \bibinfo {pages} {022330} (\bibinfo {year} {2017})}\BibitemShut {NoStop}%
\bibitem [{\citenamefont {Alexander}\ \emph {et~al.}(2020)\citenamefont {Alexander}, \citenamefont {Kanazawa}, \citenamefont {Egger}, \citenamefont {Capelluto}, \citenamefont {Wood}, \citenamefont {Javadi-Abhari},\ and\ \citenamefont {McKay}}]{alexander_qiskit_2020}%
  \BibitemOpen
  \bibfield  {author} {\bibinfo {author} {\bibfnamefont {T.}~\bibnamefont {Alexander}}, \bibinfo {author} {\bibfnamefont {N.}~\bibnamefont {Kanazawa}}, \bibinfo {author} {\bibfnamefont {D.~J.}\ \bibnamefont {Egger}}, \bibinfo {author} {\bibfnamefont {L.}~\bibnamefont {Capelluto}}, \bibinfo {author} {\bibfnamefont {C.~J.}\ \bibnamefont {Wood}}, \bibinfo {author} {\bibfnamefont {A.}~\bibnamefont {Javadi-Abhari}},\ and\ \bibinfo {author} {\bibfnamefont {D.}~\bibnamefont {McKay}},\ }\bibfield  {title} {\bibinfo {title} {Qiskit {Pulse}: {Programming} {Quantum} {Computers} {Through} the {Cloud} with {Pulses}},\ }\href {https://doi.org/10.1088/2058-9565/aba404} {\bibfield  {journal} {\bibinfo  {journal} {Quantum Science and Technology}\ }\textbf {\bibinfo {volume} {5}},\ \bibinfo {pages} {044006} (\bibinfo {year} {2020})}\BibitemShut {NoStop}%
\bibitem [{Note1()}]{Note1}%
  \BibitemOpen
  \bibinfo {note} {Https://doi.org/10.54499/QuantERA/0003/2021}\BibitemShut {NoStop}%
\bibitem [{\citenamefont {Stephens}\ \emph {et~al.}(2021)\citenamefont {Stephens}, \citenamefont {Cutshall}, \citenamefont {McPhee},\ and\ \citenamefont {Beck}}]{stephens_self-consistent_2021}%
  \BibitemOpen
  \bibfield  {author} {\bibinfo {author} {\bibfnamefont {A.}~\bibnamefont {Stephens}}, \bibinfo {author} {\bibfnamefont {J.~M.}\ \bibnamefont {Cutshall}}, \bibinfo {author} {\bibfnamefont {T.}~\bibnamefont {McPhee}},\ and\ \bibinfo {author} {\bibfnamefont {M.}~\bibnamefont {Beck}},\ }\bibfield  {title} {\bibinfo {title} {Self-consistent state and measurement tomography with fewer measurements},\ }\href {https://doi.org/10.1103/PhysRevA.104.012416} {\bibfield  {journal} {\bibinfo  {journal} {Physical Review A}\ }\textbf {\bibinfo {volume} {104}},\ \bibinfo {pages} {012416} (\bibinfo {year} {2021})}\BibitemShut {NoStop}%
\bibitem [{\citenamefont {Nielsen}\ \emph {et~al.}(2021)\citenamefont {Nielsen}, \citenamefont {Gamble}, \citenamefont {Rudinger}, \citenamefont {Scholten}, \citenamefont {Young},\ and\ \citenamefont {Blume-Kohout}}]{nielsen_gate_2021}%
  \BibitemOpen
  \bibfield  {author} {\bibinfo {author} {\bibfnamefont {E.}~\bibnamefont {Nielsen}}, \bibinfo {author} {\bibfnamefont {J.~K.}\ \bibnamefont {Gamble}}, \bibinfo {author} {\bibfnamefont {K.}~\bibnamefont {Rudinger}}, \bibinfo {author} {\bibfnamefont {T.}~\bibnamefont {Scholten}}, \bibinfo {author} {\bibfnamefont {K.}~\bibnamefont {Young}},\ and\ \bibinfo {author} {\bibfnamefont {R.}~\bibnamefont {Blume-Kohout}},\ }\bibfield  {title} {\bibinfo {title} {Gate {Set} {Tomography}},\ }\href {https://doi.org/10.22331/q-2021-10-05-557} {\bibfield  {journal} {\bibinfo  {journal} {Quantum}\ }\textbf {\bibinfo {volume} {5}},\ \bibinfo {pages} {557} (\bibinfo {year} {2021})}\BibitemShut {NoStop}%
\bibitem [{\citenamefont {Mezzadri}(2006)}]{mezzadri_how_2007}%
  \BibitemOpen
  \bibfield  {author} {\bibinfo {author} {\bibfnamefont {F.}~\bibnamefont {Mezzadri}},\ }\bibfield  {title} {\bibinfo {title} {How to generate random matrices from the classical compact groups},\ }\href@noop {} {\bibfield  {journal} {\bibinfo  {journal} {Notices of the American Mathematical Society}\ }\textbf {\bibinfo {volume} {54}} (\bibinfo {year} {2006})}\BibitemShut {NoStop}%
\end{thebibliography}%

\appendix

\section{Retrieving SPAM parameters \label{sec:SPAMTom}}
The task of self-consistent SPAM tomography consists of simultaneously estimating a density matrix, $\tilde{\rho}_0$, and 
a POVM set $\tilde{M}_l$ describing the initial state and the measuring device, respectively. Simpler versions of separate state and detector tomography 
reconstruct either the state or a POVM, assuming prior knowledge of the other \cite{eisert_quantum_2020,gebhart_learning_2023}. Nonetheless, these approaches may incur in significant errors when this knowledge is inaccurate. This frailty led to the development of self-consistent methods, in which minimal previous knowledge of state or measurement apparatus is required \cite{stephens_self-consistent_2021, nielsen_gate_2021}.
In this work, we  employ a slightly modified version of the protocol applied in \cite{wold_universal_2024}.
The backbone of the algorithm is very similar to that of the Liouvillian tomography procedure: a regression problem over Pauli string probabilities.
However, when estimating SPAM parameters, a state, $\rho_i$, is prepared and immediately measured without applying $\Lambda_t$ as represented in panel b) of Fig. \ref{fig:PauliCircuits}. Consequently, the Pauli string probabilities are now modeled by  

\begin{equation}
    p_{l|i}(t)^\text{SPAM} = \Tr{R_i\rho_0 R_i^\dagger M_{l}}.\label{eq:ProbSPAM}
\end{equation}
This model is then fitted to experimentally obtained probabilities, $\tilde{p}^{\text{SPAM}}_{l|i}$, by minimizing the MSE cost function, in order to obtain the estimates $\tilde{\rho}_0 = \rho(\tilde{\theta}_1)$ and $ \tilde{M}_l = M_l(\tilde{\theta}_2)$, where
\begin{align}
    \tilde{\theta}_1, \tilde{\theta}_2 &= \arg \min_{\theta_1, \theta_2}\times\label{eq:OptimSPAM}\\
    \times &\frac{1}{2^{N_q} N_p}\sum_{il} \left[\Tilde{p_{l|i}}^\text{SPAM} - \Tr{R_i \rho(\theta_1) R_i^\dagger M_l(\theta_2)}\right]^2.\nonumber
\end{align}
Once again, the optimization is performed over an unconstrained parameter space, $(\theta_1,\theta_2)$, which is then mapped to $(\tilde{\rho}_0, \tilde{M}_l)$.
Particularly, for the density matrix we consider the mapping 
\begin{align}
    \rho(\theta) = \frac{\theta\theta^\dagger}{\Tr{\theta\theta^\dagger}}, \quad \theta \in \mathbb{C}^{d\times d},
\end{align}
which guarantees positivity and unit trace. 
For the POVM parameterization, the Kraus operator mapping of Eq. \ref{eq:MapParam} is employed with $r = d$,
\begin{align}
    M_l(\theta) = E_l^\dagger(\theta) E_l(\theta),\quad \theta\in\mathbb{C}^{d^2 \times d}.
    \label{eq:ParamPOVM}
\end{align}
The multiplication by the adjoint ensures each POVM operator will be positive and, together with the trace-preservation condition of $E_l$, guarantees that the condition $\sum_l M_l = 1$ is satisfied.

The optimization task of Eq. \ref{eq:OptimSPAM} is solved numerically with the Adam optimizer.

To benchmark SPAM retrieval, we consider SPAM target sets consisting of perturbations to the ideal 
pure state $\ket{0}\bra{0}$ and the projective operators $\ket{l}\bra{l}$ over the computational basis, 
\begin{subequations}
\begin{align}
    \rho_0  &= 0.9\ket{0}\bra{0} + 0.1\delta \rho, \\
    M_l &= 0.8\ket{l}\bra{l} + 0.2 \delta M_l.\label{eq:PertPOVM}
\end{align}
\label{eq:PertSPAM}
\end{subequations}
The perturbations $\delta\rho,\, \delta M_l$ are obtained by computing $\rho(A_1+iB_1)$ and $M_l(A_2+iB_2)$, where 
$A_1,B_1 \in \mathbb{R}^{d\times d},\, A_2, B_2 \in \mathbb{R}^{d^2 \times d}$ with each matrix entry sampled independently 
from a standardized normal distribution. The coefficients of the perturbations reflect previously obtained values in superconducting devices \cite{samach_lindblad_2022}.

From a target SPAM set, Pauli string probabilities can be obtained by Eq. \ref{eq:ProbSPAM}. Using these probabilities 
directly would correspond to $N_s = \infty$, therefore to include finite sampling effects,
the probabilities $p_{l|i}^\text{SPAM}$ are sampled $N_s$ times and $\tilde{p}_{l|i}^\text{SPAM}$ are approximated by the relative frequencies. 
The target set is compared to the retrieved SPAM set by computing the state and POVM fidelities of Eq. \ref{eq:Fidelity},

\begin{figure}
    \includegraphics[width=0.45\textwidth]{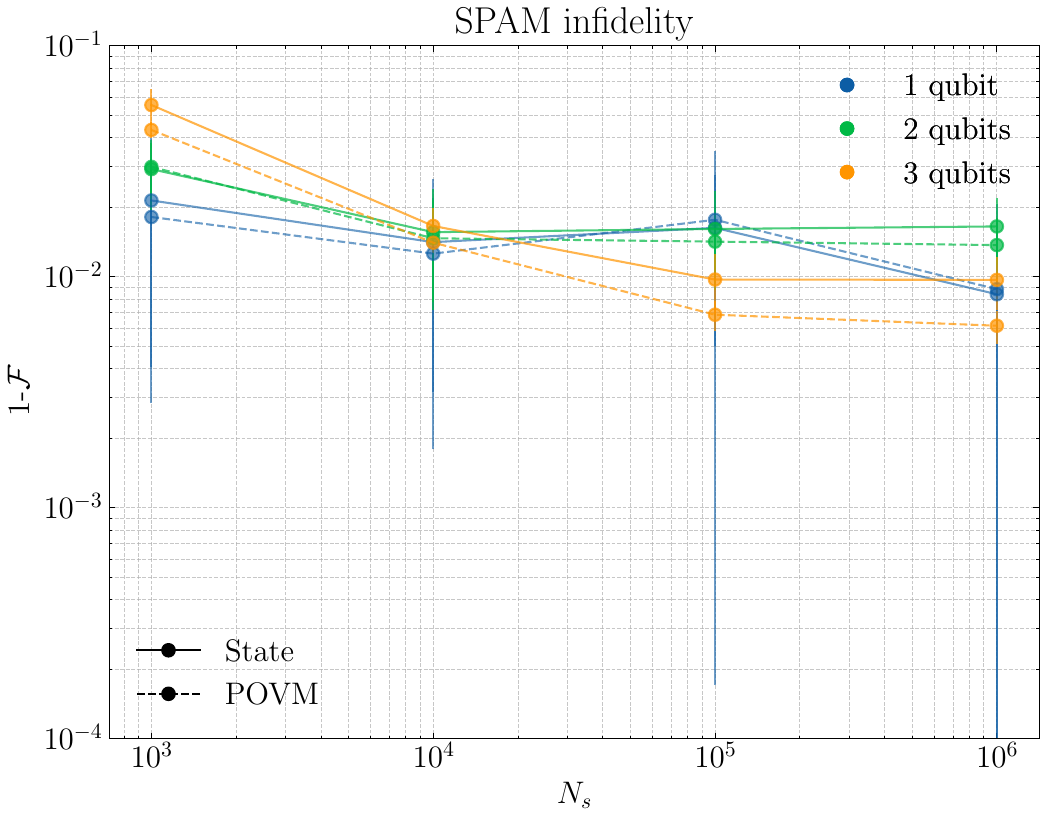}
    \caption{Benchmark results for SPAM tomography: Average state and POVM infidelity between target and reconstructed SPAM set as a function of $N_s$. The results display the mean value over an ensemble of 20 reconstructions. The error bars correspond to the standard deviation.}
    \label{fig:SPAMShots}
\end{figure}

The impact of $N_s$ on SPAM retrieval is investigated by generating 20 target SPAM sets with the corresponding 
probabilities computed with different values of $N_s$. The infidelities, $1-\mathcal{F}$, resulting from the application of
SPAM tomography to these probabilities are displayed in Fig. \ref{fig:SPAMShots}.
For one qubit the infidelity is approximately constant for all values of $N_s$, while for 2 and 3 qubits there is a decrease in infidelity until it 
eventually plateaus. As such, for all systems considered, there is a point after which further reducing the statistical noise 
no longer improves the quality of the retrieved state and POVM. This is a consequence of 
distinct SPAM sets producing the same Pauli string probabilities. Hence, each of these sets will 
correspond to equivalent MSE minima to which the optimization of Eq. \ref{eq:OptimSPAM} may converge. 
As such, the optimization problem and, consequently, the SPAM tomography protocol is ill-posed.
This ill-posed nature may be restricted by replacing Pauli strings by more general probes such as 
Haar unitary circuits. However, there will be a lingering ambiguity in retrieving the SPAM set 
which is intrinsic to self-consistent methods \cite{stephens_self-consistent_2021,nielsen_gate_2021}. To hinder these effects, a bias is introduced during optimization by setting the initial position to be in the vicinity of the ideal SPAM parameters.  

Despite the aforementioned limitations, the proposed SPAM tomography algorithm is capable of retrieving high-fidelity estimates sufficient for accurate Liouvillian tomography.

\section{Retrieving CPTP maps \label{sec:MapTom}}

Consider the set-up of section \ref{sec:PS}, process tomography aims at reconstructing an unknown quantum operation, $\Lambda_t$, in the form of a CPTP map.

A CPTP map supports several representations, which may be leveraged for different tomographic schemes, with many algorithms having been proposed \cite{eisert_quantum_2020,gebhart_learning_2023}. 
In this work, we present the algorithm introduced in \cite{wold_universal_2024}, in which the Kraus or operator-sum representation is used.
In this representation the CPTP map is fully described by a set of operators, $\{E_\mu\}$, satisfying the trace-preservation relation $\sum_{\mu = 1}^r E_\mu^\dagger E_\mu = 1$, where $r$ is the Kraus rank of the map \cite{watrous_theory_2018}.  
The action of the map on a state is then given by $\Lambda (\rho)  = \sum_{\mu = 1}^r E_\mu \rho E_\mu^\dagger$. 
Note that a single CPTP map can be described by several sets of operators, possibly with differing Kraus ranks. These sets are connected by unitary transformations 
$E_\nu' = \sum_\mu u_{\nu\mu}E_\mu$, with $u u^\dagger = 1$.

As was suggested in section \ref{sec:PS}, the Pauli string distributions provide a way of estimating the CPTP map. 
Similarly to Liouvillian tomography, the model of Eq. \ref{eq:PauliStringProb} is fitted to the experimentally obtained probabilities $\tilde{p}_{l|ij}(\tau)$, 
yielding $r$ operators $\Tilde{E}_\mu = E_\mu(\Tilde{\theta})$, where  

\begin{align}
    \tilde{\theta} &= \arg\min_{\theta} \frac{1}{2^{N_q} N_p} \times \label{eq:QPTOptimiConstrained}\\
    &\times \sum_{ijl}\left[\Tilde{p}_{l|ij}(\tau)- \Tr{ \sum_\mu E_\mu(\theta) \rho_i E_\mu^\dagger (\theta) M_{lj}}\right]^2.\nonumber
\end{align}
Once again, the optimization is done over the space of unconstrained parameters $\theta \in \mathbb{C}^{rd\times d}$, which are mapped to a CPTP map 
through the parameterization

\begin{subequations}
\label{eq:MapParam}
\begin{align}
    Q,R &= \text{QR}(\theta),\\
    D_{ij} &=\delta_{ij}\frac{R_{ii}}{|R_{ii}|}\\
    U &= QD,\\
    E_\mu &= U[\mu:\mu+d-1],\quad 1 \leq \mu \leq r.
\end{align}  
\end{subequations}

In this way, when generating a map with Kraus rank $r$, the QR-decomposition is applied to unconstrained parameters $\theta$ to produce a semi-unitary
matrix $U$, which is then sliced row-wise to obtain $r$ Kraus operators defining the Kraus representation of a CPTP map. The unitarity of $U$ ensures that 
the trace-preservation condition is satisfied. Nonetheless, the QR-decomposition of a matrix is not unique as it is invariant to transformations
$Q' = QD,\, R' = D^{-1}R$, for all unitary diagonal $D\in\mathbb{C}^{d\times d}$ \cite{mezzadri_how_2007}.

This ambiguity results in the same parameterization producing two distinct maps, corresponding to different choices of $D$.
Thus, when performing numerical optimization, a small perturbation in parameter space could imply a large perturbation on the Kraus operators, which would significantly 
thwart optimization. 
Therefore, $D$ is fixed by forcing the diagonal elements of $R$ to be real and positive. In this convention, $D_{ij} = \delta_{ij}\frac{R_{ii}}{|R_{ii}|}$,
is the diagonal matrix obtained by dividing the diagonal entries of $R$ by their norm. 

Despite fixing the ambiguity of the QR decomposition, this parameterization is still many-to-one, as different choices of $\theta$ may lead to 
the same Kraus operators. As an example, consider parameters $\theta_1\in \mathbb{C}^{rd\times d}$ corresponding to
an intermediate semi-unitary matrix $U_{\theta_1}$ and a set of Kraus Operators $E_\mu$.
The parameters $\theta_2 = U_{\theta_1}$ will also correspond to the same set of Kraus operators, $E_\mu$, since the $\theta_2$ is already semi-unitary. 
Moreover, different parameters can produce distinct Kraus operators which are just different representations of the 
same CPTP map due to the map's invariance to unitary transformations of the Kraus operators.
 Consequently, there will be many equivalent minima for the loss, resulting in a non-convex optimization problem, which is tackled with the 
aforementioned Adam optimizer \cite{kingma2017adammethodstochasticoptimization}.

To benchmark process tomography, synthetic data is generated by first sampling a random Kraus map: For a chosen rank 
$r$, two matrices, $A,B\in \mathbb{C}^{rd\times d}$, are obtained by sampling each entry independently from a standardized normal distribution. 
The Kraus operators describing the CPTP map are then computed as $E_\mu (A+iB),$ for $1\leq \mu\leq r$.
Using the sampled CPTP map, $\{E_\mu\}$, a target SPAM set obtained by Eq. \ref{eq:PertSPAM}, and Eq. \ref{eq:PauliStringProb}, synthetic Pauli string probabilities can be generated.
Once more, to mimic experimental conditions the distributions $p_{l|ij}$ are sampled $N_s$ times and the relative frequencies are taken. 

In order to assess the influence of the number of shots, $N_s$, in the model's performance, twenty full-rank maps were generated and retrieved 
using Pauli string probabilities obtained with differing numbers of shots for 1,2 and 3 qubits. The SPAM parameters were assumed to be known exactly. 

\begin{figure}
    \subfigure{\includegraphics[width=0.45\textwidth]{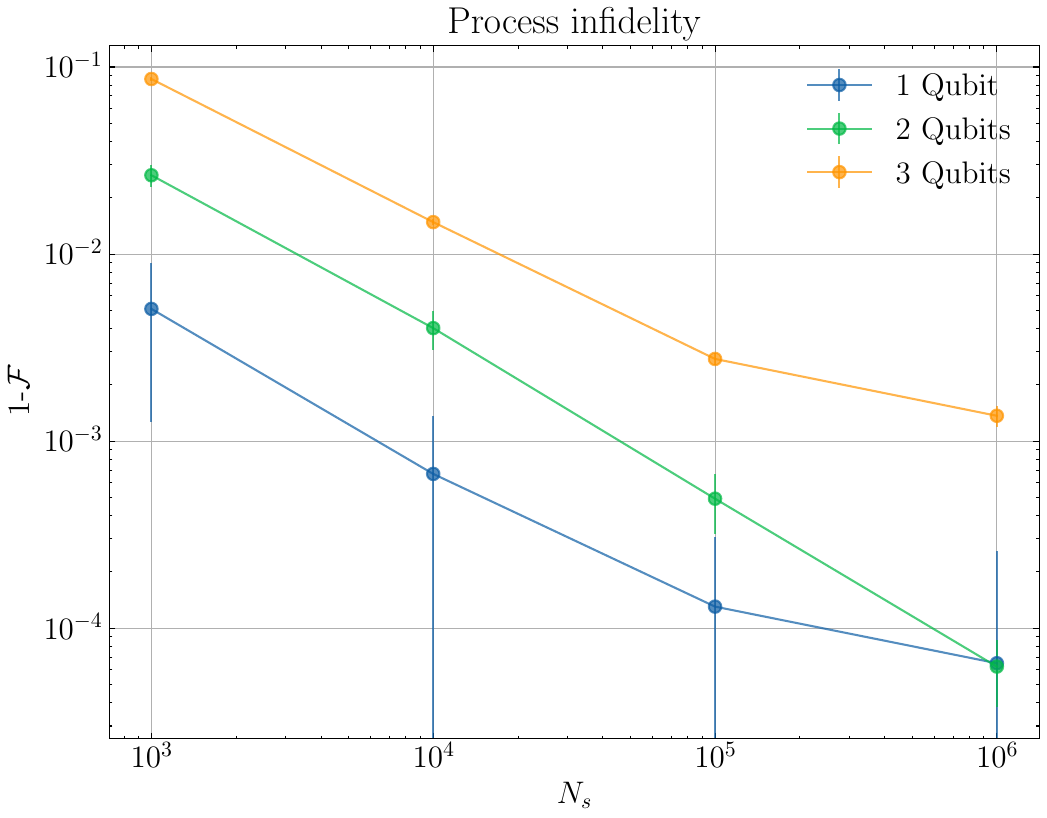}}  
    \caption{Benchmark results for process tomography: process infidelity  between target and reconstructed CPTP maps as a function of $N_s$. The results display the mean value over an ensemble of 20 reconstructions. The error bars correspond to the standard deviation.}
    \label{fig:QPTFidShots}
\end{figure}

The retrieved maps were compared to the target by computing the process infidelity between the two, $1-\mathcal{F}(\Lambda,\tilde{\Lambda})$. The results are shown in Fig. \ref{fig:QPTFidShots}.

With the increase in the number of shots, the infidelity lowers for the three systems sizes considered, 
reflecting the impact of more precise knowledge of the Pauli string distributions on the algorithm's performance. 
Unlike for SPAM tomography, the infidelity does not appear to stabilize, suggesting 
that, in the absence of noise, the QPT protocol should be capable of retrieving the CPTP map exactly.
In this sense, despite the existence of multiple equivalent minima of the loss function over the space of 
unconstrained parameters, these are expected to all correspond to the same physical CPTP map.
Hence, the optimization of Eq. \ref{eq:QPTOptimiConstrained} appears to be a physically well-posed problem.

The results presented in Fig. \ref{fig:QPTFidShots} assume perfect knowledge of the SPAM parameters. 
If these are estimated via self-consistent SPAM tomography, they will be affected by errors that will 
be propagated to the retrieved CPTP map. However, on average, the achieved fidelities were found to be only marginally inferior. 

\section{Sampling non-Markovian dynamics \label{eq:SampNonMarkov}}

To sample general non-Markovian dynamics we consider a system of $N_q = 2$ qubits interacting with an environment of $N_E = 2$ qubits which, in turn, are connected to a Markovian reservoir. 
The joint system and environment dynamics are described by a static Liouvillian in the vectorized representation with jump operators acting trivially on the system's degrees of freedom. 
\begin{subequations}
\begin{align}
    &\mathcal{L}_{\mathcal{S}+\mathcal{E}} = -i\left(H\otimes \mathds{1} - \mathds{1}\otimes H^T \right) +\label{eq:ReservoirLiouvillian}\\
    &+\frac{\alpha}{4^{N_q}-1} \sum_{\mu=1}^{4^N_E - 1} J_\mu \otimes J_\mu^* -\frac{1}{2}\left(J_\mu^\dagger J_\mu \otimes \mathds{1} + \mathds{1} \otimes \left(J_\mu^\dagger J_\mu\right)^T \right),\nonumber\\
    H &= H_{\mathcal{S}}\otimes \mathds{1}_{\mathcal{E}} + \mathds{1}_\mathcal{S}\otimes H_{\mathcal{E}} + 2gH_{\text{Int}}\label{eq:ReservoirHamil},\\
    J_\mu &= \mathds{1}_\mathcal{S}\otimes J_{\mathcal{E}\mu},\label{eq:ReservoirJOp}
\end{align}
\end{subequations}
where the Hamiltonians, $H_{\mathcal{S}},\,H_{\mathcal{E}},\, H_{\text{int}}$, are sampled from the Gaussian unitary ensemble,
and each jump operator is taken as $j_{\mathcal{E}\mu} = A+iB$, where the entries of $A,B\in\mathbb{R}^{2^{N_E} \times 2^{N_E}}$ are sampled 
independently from a Gaussian distribution. The coupling constants were chosen to be $\alpha = 1,\, g = 0.5$. It has been suggested that for a large enough environment, $\mathcal{E}$, this setup captures general non-Markovian phenomena \cite{chruscinski_non-markovian_2010}.

Assuming an initial joint system and environment product state $\ket{\ket{\rho_0\otimes\rho_{\mathcal{E}}}} = \sum_{jj'nn'}v^s_{jj'}v^{\mathcal{E}}_{'nn'}\ket{\ket{jnj'n'}}$, the 
vectorized representation of the CPTP map describing system's evolution is obtained by taking the exponent of the Liouvillian and 
tracing over the environmental degrees of freedom
\begin{subequations}
\label{eq:LambdaStein}
\begin{align}
    \rho(t) &= \Lambda_t\rho_0\\
    \bra{ii'}\Lambda_t\ket{jj'} = \sum_{mnn'}&\bra{\bra{im i'm}}e^{t\mathcal{L}^{\mathcal{S}\cup\mathcal{E}}}\ket{\ket{jnj'n'}} v^{\mathcal{E}}_{nn'}
\end{align}
\end{subequations}

\end{document}